\documentclass[aps,twocolumn,eps,epsf,floats,pre,showpacs,floatfix]{revtex4-1}
\usepackage{graphicx} 
\usepackage{amsmath}
\usepackage{color}

\newcommand{\p}{\partial}

\newcommand \be {\begin{equation}}
\newcommand \ee {\end{equation}}
\newcommand \bea {\begin{eqnarray}}
\newcommand \eea {\end{eqnarray}}

\newcommand \ppi{\mbox{\boldmath$\pi$}}

\newcommand \eps{\epsilon}
\newcommand \cali {{\cal I}}

\newcommand \lx {\left}
\newcommand \rx {\right}

\newcommand \cH {{\cal H}}
\newcommand \cI {{\cal I}}
\newcommand \cB {{\cal B}}
\newcommand \cZ {{\cal Z}}

\newcommand \sumI {\sum_{{\rm i}\in \cI}}

\newcommand \sumJ {\sum_{{\rm j}\in \cI}}

\newcommand \sumIJ {\sum_{{\rm i,j}\in \cI}}

\newcommand \sumA {\sum_{{\rm a}\in \cB}}

\newcommand \sumAB {\sum_{{\rm a,b}\in \cB}}

\newcommand \Zt {\cZ\lx(J,g,\mu\rx)}

\newcommand \Ei {\epsilon_{\rm i}}
\newcommand \Ej {\epsilon_{\rm j}}
\newcommand \Ea {\epsilon_{\rm a}}

\newcommand \Paia {\vec{\mathbf{\pi}}_{\rm a}}

\newcommand \Hi {\vec{\mathbf{h}}_{\rm i}}
\newcommand \Hj {\vec{\mathbf{h}}_{\rm j}}

\newcommand \Vi {\vec{\mathbf{v}}_{\rm i}}
\newcommand \Vj {\vec{\mathbf{v}}_{\rm j}}
\newcommand \Va {\vec{\mathbf{v}}_{\rm a}}
\newcommand \Vb {\vec{\mathbf{v}}_{\rm b}}

\begin{document}

\title{Social interactions dominate speed control in driving natural flocks toward criticality} 

\author{William Bialek$^a$}
\author{Andrea Cavagna$^b$}
\author{Irene Giardina$^b$}
\author{Thierry Mora$^c$}
\author{Oliver Pohl$^b$}
\thanks{Present address:  Institut f\"ur Theoretische Physik, Technische Universit\"at Berlin, Hardenbergstrasse 36, D--10623 Berlin--Charlottenburg, Germany.}
\author{Edmondo Silvestri$^b$}
\author{Massimiliano Viale$^b$}
\author{Aleksandra Walczak$^d$}

\affiliation{$^a$Joseph Henry Laboratories of Physics and Lewis--Sigler Institute for Integrative Genomics, Princeton University, Princeton, New Jersey 08544 USA\\
$^b$Istituto Sistemi Complessi (ISC--CNR), Via dei Taurini 19, 00185 Roma, Italy\\
$^b$Dipartimento di Fisica, ``Sapienza'' Universit\'a di Roma,  P.le Aldo Moro 2, 00185 Roma, Italy\\
$^c$Laboratoire de Physique Statistique de lÕ'{\'{E}cole} Normale Sup\'erieure, CNRS and Universites Paris VI and Paris VII, 24 rue Lhomond, 75231 Paris Cedex 05, France,  and\\
$^d$Laboratoire de Physique Th\'eorique de lÕ'\'Ecole Normale Sup\'erieure, CNRS and University Paris VI, 24 rue Lhomond, 75231 Paris Cedex 05, France}

\date{\today} 

\begin{abstract}
Flocks of birds exhibit a remarkable degree of coordination and collective response.  It is not just that thousands of individuals fly, on average, in the same direction and at the same speed, but that even the fluctuations around the mean velocity are correlated over long distances.  Quantitative measurements on flocks of starlings, in particular, show that these fluctuations are scale--free, with effective correlation lengths proportional to the linear size of the flock.  Here we construct models for the joint distribution of velocities in the flock that reproduce the observed local correlations between individuals and their neighbors, as well as the variance of flight speeds across individuals, but otherwise have as little structure as possible.  These minimally structured, or maximum entropy models provide quantitative, parameter--free predictions for the spread of correlations throughout the flock, and these are in excellent agreement with the data.  These models are mathematically equivalent to statistical physics models for ordering in magnets, and the correct prediction of scale--free correlations arises because the parameters---completely determined by the data---are in the critical regime.  In biological terms, criticality allows the flock to achieve maximal correlation across long distances with limited speed fluctuations.
\end{abstract}

\maketitle

\section{Introduction}

In a flock of birds, thousands of individuals will fly  in the same direction and at the same speed, for long periods of time.  But this average behavior  is not enough for flocking to be advantageous.  The entire flock must  respond to dangers that may be visible only to a small fraction of individuals, requiring information to propagate over long distances.  Although it is difficult to measure this information flow directly \cite{attanasi+al_13}, we know that attacks by predators on a flock have very low success rates \cite{pulliam_73,cresswell_94,krause+ruxton_02}, and that the evasion of predators by starling flocks is associated with the triggering and propagation of waves through  the flock \cite{procaccini+al_11}.  Even in the absence of predators, we can see deviations of individual behavior from the average behavior of the flock, and correlations in these fluctuations provide a signature of information flow through the flock.  Strikingly, observations on flocks of starlings show that these correlations extend over very long distances, comparable to the size of the flock itself \cite{cavagna+al_10}.

It is generally believed that the interactions among birds in a flock are local---each bird  aligns its flight direction and speed to those of its near neighbors \cite{couzin+krause_03}.  If this is correct, then we have to understand how local interactions can generate correlations over much longer distances.  In physics, we have two very different mechanisms for local interactions to produce correlations that are essentially scale--free, extending over distances comparable to the size of the system as a whole.  If the system spontaneously breaks a continuous symmetry, for example when all the spins in a magnet select a particular direction in space along which the macroscopic magnetization will point, then the fluctuations in the system are dominated by ``Goldstone modes'' that do not decay on any fixed length scale \cite{stat-mech}.  If we can think of the alignment of flight directions in a flock as being like the alignment of spins in a magnet \cite{toner+tu_95,toner+tu_98,ramaswamy_10}, then we can understand the emergence of scale--free correlations by analogy with Goldstone's theorem.  We have shown that this is more than a metaphor \cite{bialek+al_12}: the minimally structured model consistent with the observed correlations among flight directions of neighboring birds is exactly equivalent to a model of spins in a magnet, and the resulting (parameter--free) prediction of long ranged correlations among fluctuations in flight direction agrees quantitatively with the data.

Not just the fluctuations in flight direction, but also the fluctuations in flight speed are correlated over long distances \cite{cavagna+al_10}.  Now there is no  analogy to Goldstone modes, because choosing a speed does not correspond to breaking any plausible symmetry of the system.  But there is a second mechanism by which physical systems generate scale--free correlations, and this is by tuning parameters to a critical point \cite{stat-mech,wilson_79}.  As we explore the parameter space of a system (e.g., changing  temperature and pressure), we encounter phase transitions, where small changes in parameters produce qualitative changes in behavior of a macroscopic sample (e.g., between liquid and gas).  Along the lines in parameter space where these phase transitions exist, there are special points, called critical points, where the dependence on parameters becomes, for very large systems, singular but not discontinuous.  At these points, fluctuations (e.g., in the density of the liquid) become correlated on all length scales, from the molecular scale of the interactions to the macroscopic scale of the sample as a whole.  

Tuning to a critical point provides a potential explanation for the observed scale--free correlations in speed of flocking birds, but this is just an analogy;   the goal of this paper is to see if we can construct a quantitative theory.  Our strategy follows that in Ref \cite{bialek+al_12}:  we will construct the least structured models that are consistent with measured correlations among neighboring birds, and then see if these models can predict correctly the persistence of correlations over much longer distances, comparable to the size of the flock.  We will see that this does work, and that the underlying mechanism really is the tuning of the system to a critical point.  From a biological point of view, this tuning means that individuals in a flock combine individual speed control and social interactions with their neighbors so as to achieve a maximal range of influence while keeping speed variability low. 

\section{Building a model  from  data}
\label{building}

We consider flocks of European starlings, {\it Sturnus vulgaris}, in the field.   The work of Refs \cite{cavagna+al_08a,cavagna+al_08b} provides a detailed description of these flocks, resulting in the assignment of three--dimensional positions and velocities, at each moment in time, to each individual bird in flocks with up to several thousand members; for a summary see Appendix \ref{app:data}.  From these raw data, one can extract a variety of features that serve to characterize the nature of the ordering in the flock \cite{ballerini+al_08a,cavagna+al_10}, including the scale--free correlations noted above.  

The positions and velocities of all the birds in the flock are stochastic---with elements of randomness, but correlated.  In making a model, we want to be able to predict the probability distribution out of which these random variables are drawn.  One approach is to consider a detailed model for the dynamics of the flock, typically with many parameters to describe the interactions that cause the flock to cohere and align.  In this approach, the connection between the model dynamics and the joint distribution of velocities in the flock can be complicated, and fitting the parameters of the interactions is difficult.  Alternatively, we can take some set of observations on the flock as given and try to construct models that reproduce these observations exactly; among the (generally infinite) set of models that can do this, we want to choose the one that has the least structure.  Minimizing structure means that the velocities we choose out of the distribution are as random as they can be while still matching the properties of the flock that we have chosen as essential.  As emphasized by Jaynes \cite{jaynes_57,mackay}, these minimally structured distributions have maximum entropy, and this provides a connection to the ideas of statistical physics, even though we are describing a system that is not in thermal equilibrium (see Appendix \ref{maxentapp}).

The maximum entropy approach to model building is far from new, but there has been a resurgence of interest in the use of these ideas to describe biological systems  \cite{schneidman+al_06,shlens+al_06,psu_group,tang+al_08,weigt+al_09,halabi+al_09,mora+al_10,stephens+bialek_10,tkacik+al_13a,tkacik+al_13b}.  In Ref \cite{bialek+al_12}, we took a first step toward a maximum entropy description of flocks, building models for the distribution of flight directions that match the average local correlation between the direction of a bird and its nearest neighbors.    Surprisingly, fixing this one number leads to a model that, with no free parameters, provides an essentially complete, quantitative description of the propagation of directional order throughout the entire flock.  Here we want to generalize this approach to consider not just flight directions, but also speed.  As explained above, we expect that accounting for the observed properties of speed ordering is a qualitatively different problem from the case of directional ordering.

Given the positions of the birds in space, the state of the flock is defined by the velocity $\vec{\mathbf{v}}_{\rm i}$ of each bird.  This three dimensional vector is composed of the speed, $v_{\rm i} \equiv | \vec{\mathbf{v}}_{\rm i}|$, and a unit vector, $\vec{\mathbf{s}}_{\rm i} = \vec{\mathbf{v}}_{\rm i}/v_{\rm i}$, that points in the direction of flight.  Our intuition is that the most important interactions are local, between a bird and its immediate neighbors.  If this is correct, then the essential features of the system should be captured by measuring local correlations, as in Ref \cite{bialek+al_12}.  

We can quantify local correlations in the flock by asking how similar, on average, the velocity of each bird is to its neighbors.  To do this, we define 
\begin{equation}
Q_{\rm int} = {1\over {2v_0^2 N}}\sum_{{\rm i}=1}^N {1\over {n_c}}\sum_{{\rm j} \in {\cal N}_{\rm i}} |\vec{\mathbf{v}}_{\rm i} - \vec{\mathbf{v}}_{\rm j} |^2 .
\label{Qint}
\end{equation}
Here ${\cal N}_{\rm i}$ is the relevant neighborhood of bird $\rm i$, which we take to be its first $n_c$ nearest neighbors \cite{ballerini+al_08a,bialek+al_12}.  We compare a bird to each of its neighbors, average over the neighborhood, and then average over all $N$ birds in the flock; we normalize the result by a typical speed $v_0$ so that we have a dimensionless measure of correlation or similarity.   If we take $v_0$ to be the average speed of birds in the flock, then typical values for $Q_{\rm int}$ are $\sim 10^{-2}$ (Table \ref{table} in Appendix \ref{app:data}), showing that individual birds indeed fly with velocities that are very similar to those of their neighbors.

\begin{figure*}[t] 
\centering
\includegraphics[scale=0.19]{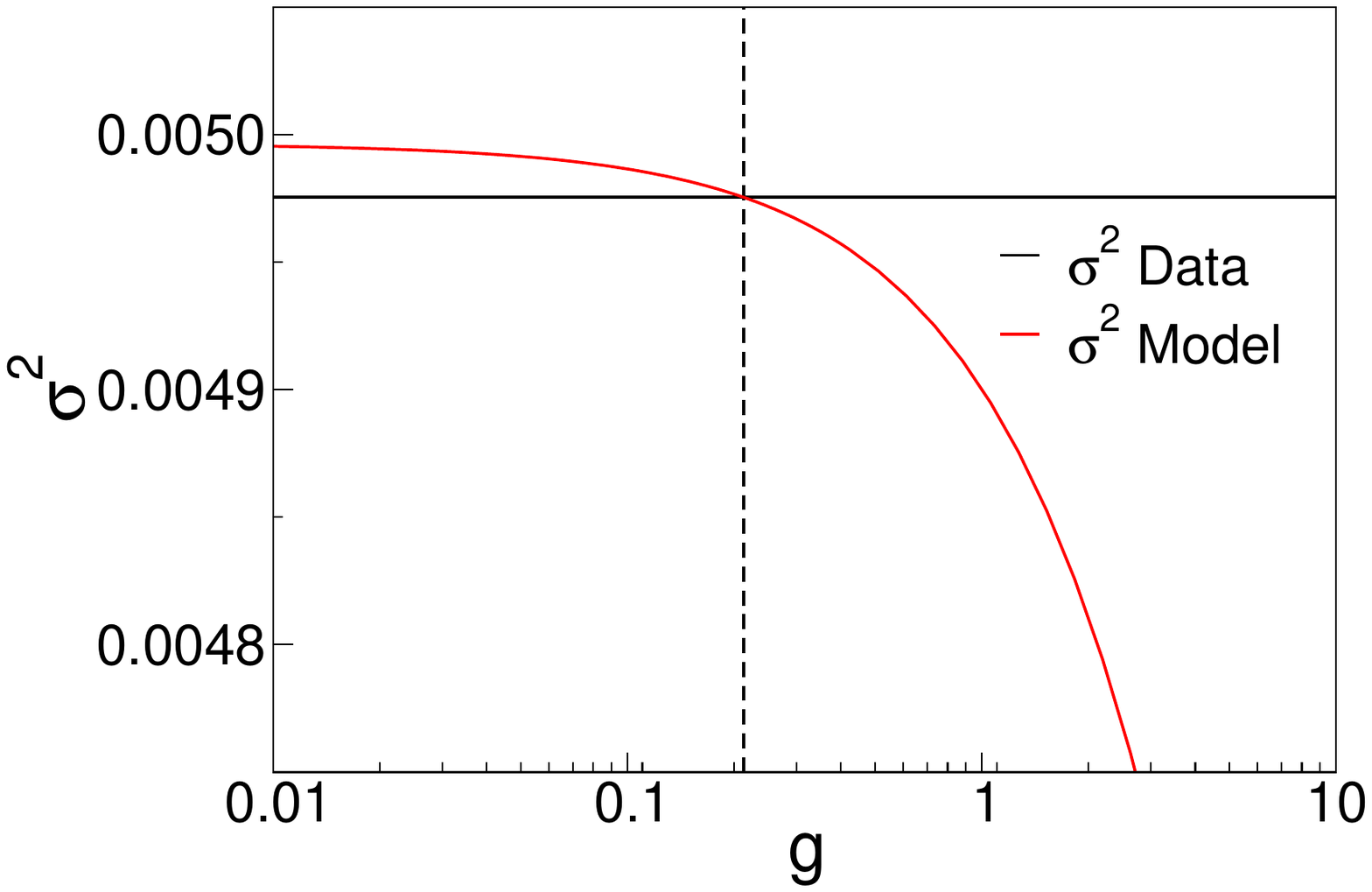}
\includegraphics[scale=0.19]{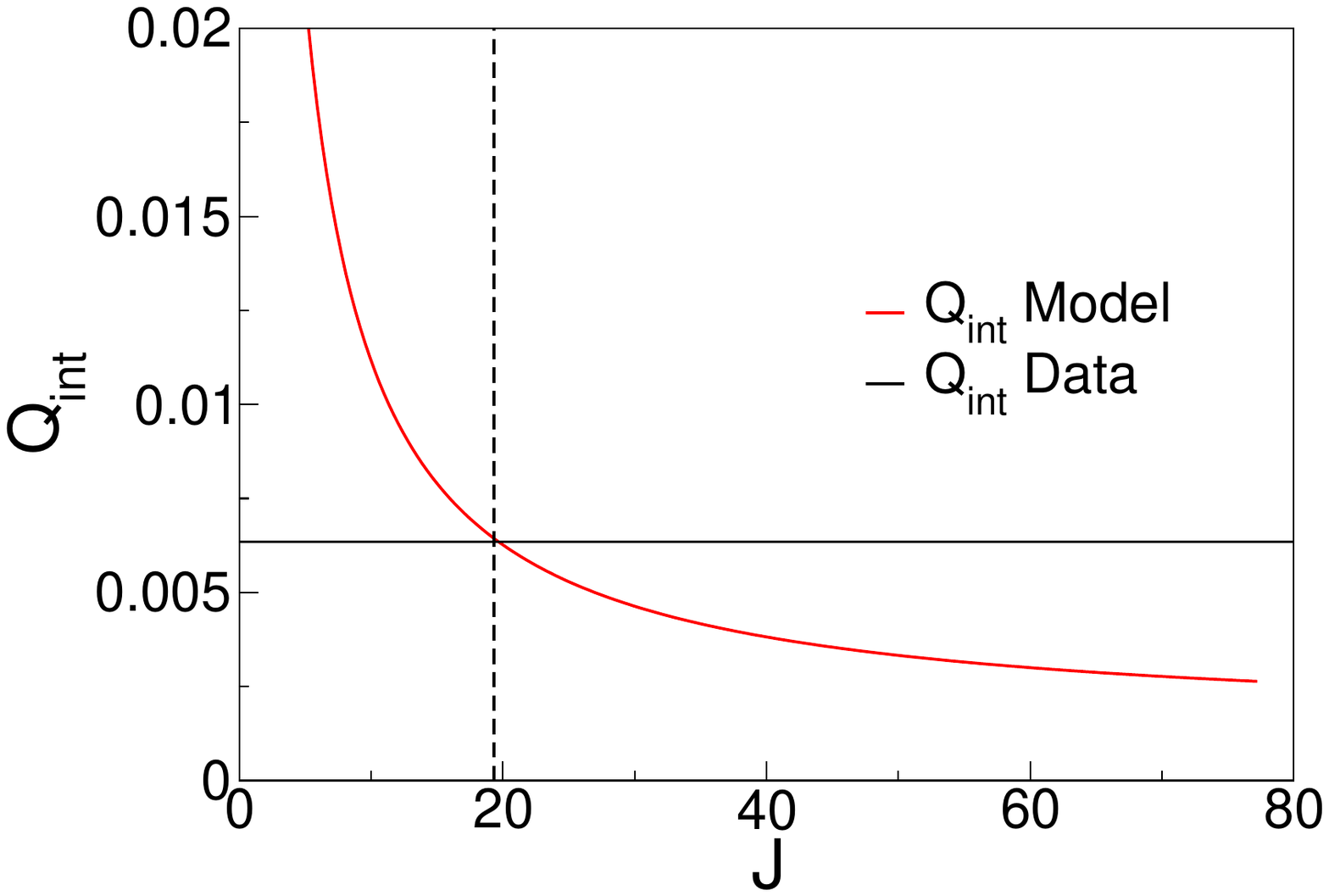} 
 \includegraphics[scale=0.19]{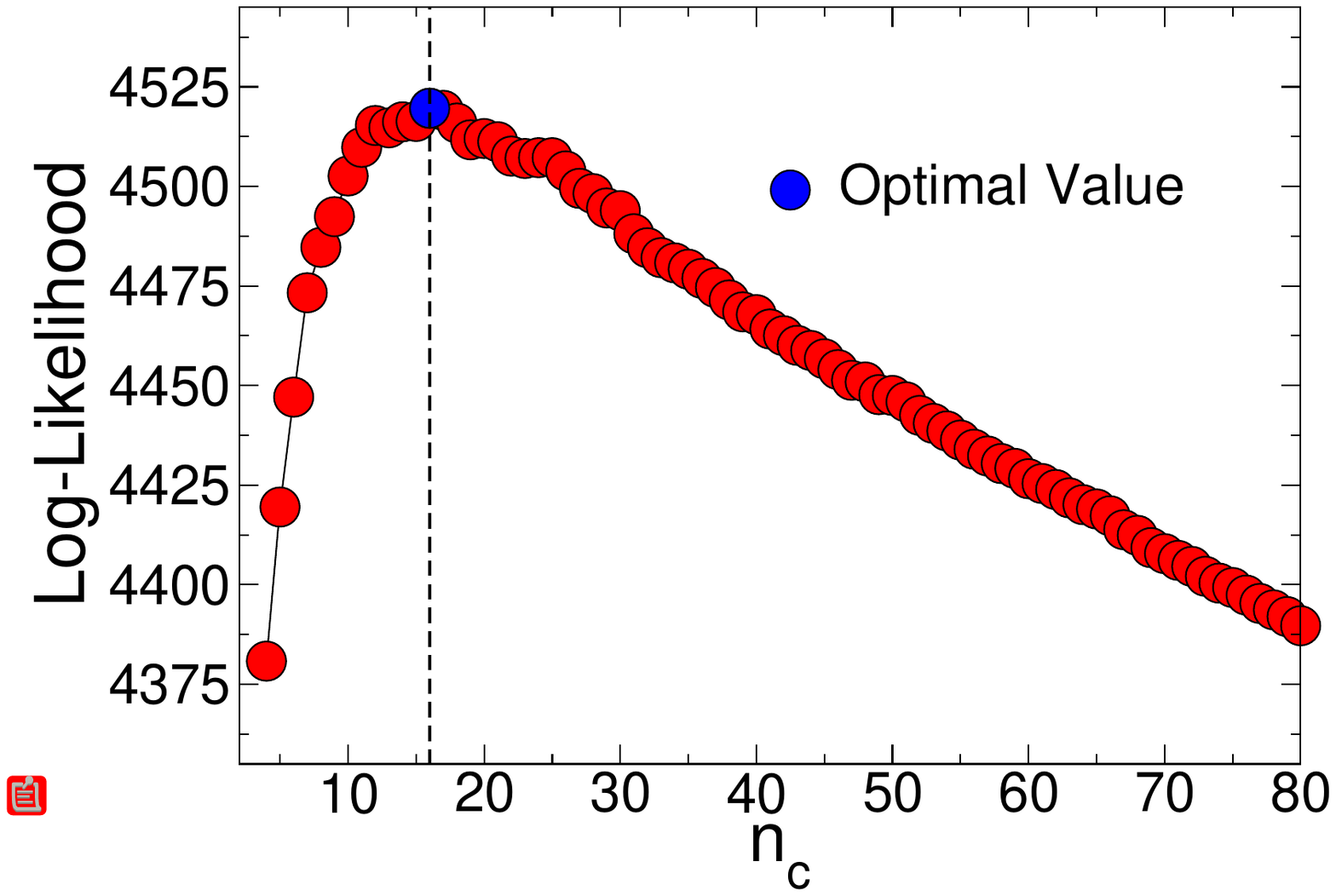}\\
 \quad\mbox{   (a)}\hspace{5.5 cm}\mbox{(b)}\hspace{5.5 cm}\mbox{(c)}
\caption{Inference of the three interaction parameters $g$,$J$ and $n_c$.
 (a) For fixed values of  $J$ and $n_c$, the value of the speed control parameter $g$ is  found by equating the theoretical prediction for the variance of fractional speed fluctuations, $\sigma^2$ (red line), to its experimental value (black horizontal line). (b) Once the value of $g$ is determined for all possible values of $J$ and $n_c$, the interaction strength $J$ can be set by equating the theoretical prediction for $Q_{\rm int}$ (red line) to its experimental value (black line). (c) Once $g$ and $J$ are computed for given values of $n_c$, the log--likelihood of the data, $\langle \ln P(\{\vec{\bf v}_{\rm i}\})\rangle_{\rm exp}$ becomes a function of $n_c$ only, and the interaction range $n_c$ can be evaluated by maximizing this function. All panels refer to the same  snapshot (image 2) of flock 25-10, and mathematical details can be found in Appendix \ref{app:solving}.   \label{fig:parameters}}
\end{figure*}

The definition of $Q_{\rm int}$ quantifies the similarity of each bird's flight vector to that of its neighbors, but it can't completely specify the structure of the flock.  If we add a constant to all the velocities, so that the flock flies faster or slower, then $Q_{\rm int}$ is unchanged.  We would like to match the average speed of the birds in the flock, $V = {1\over N} \sum_{{\rm i}=1}^N v_{\rm i} $, to its observed value $\langle V\rangle_{\rm exp}$. In addition, we know that individual birds have speeds that vary around the mean, so we would also like to match the variance of speeds.  This is equivalent to fixing not just the mean speed, but the mean square speed,  $V_2 = {1\over N} \sum_{{\rm i}=1}^N v_{\rm i}^2$.

The maximum entropy distribution consistent with measured values of $Q_{\rm int}$, $V$, and $V_2$ has the form (see Appendix B),
\bea
P(\{\vec{\mathbf{v}}_{\rm i}\}) &=&\frac{1}{Z}\exp\left[-\frac{J}{4v_0^2} \sum_{{\rm ij}=1}^N n_{\rm ij} 
|\vec{\mathbf{v}}_{\rm i}-\vec{\mathbf{v}}_{\rm j}|^2 \right. \nonumber\\
&&\,\,\,\,\,\,\,\,\,\,\,\,\,\,\,\,\,\,\,\,\,\,\,\,\,\,\,\,\,\, 
\left. +{\mu \over {v_0}} \sum_{{\rm i}=1}^N v_{\rm i}-\frac{g}{2v_0^2}\sum_{{\rm i}=1}^N v_{\rm i}^2 \right],
\label{maxent}
\eea
where $Z$ is a constant that ensures the normalization of the probability distribution, and we have inserted  factors of $v_0$ so that  other parameters are dimensionless.  The matrix $n_{\rm ij}$ maps the connections between birds:  $\hat n_{\rm ij} = 1$ if bird j  is in the neighborhood of bird ${\rm i}$ (${\rm j}\in {\cal N}_{\rm i}$), and zero otherwise; we symmetrize to give $n_{\rm ij} = (\hat n_{\rm ij} + \hat n_{\rm ji})/2$.  The parameters $J$, $\mu$, and $g$ must be adjusted so that the average values of $Q_{\rm int}$, $V$, and $V_2$ computed from the probability distribution match those observed for the flock; as explained in Appendix \ref{app:solving}, these computations can be done analytically.  The only remaining parameter is the number of relevant neighbors $n_c$, which we fix by requiring that the probability of the observed velocities be as large as possible (maximum likelihood).  

Figure \ref{fig:parameters} shows one example of our solution to the ``inverse problem'' of determining the parameters $J$, $g$, and $n_c$.  Importantly, the quantities that we are trying to match are averages over all the birds in the flock, and so they are determined with small errors even from a single snapshot of the velocities.  The parameters in turn are determined very precisely, and are consistent for a single flock across time, as in Ref \cite{bialek+al_12}.  

\section{Some intuition}
\label{sec:intuition}

Maximum entropy distributions are mathematically equivalent to the Boltzmann distribution for systems in thermal equilibrium, and we can use this identity to gain some intuition for the predictions of the model.   We recall that a system described by the Boltzmann distribution will occupy a state $s$ with probability $p_s \propto \exp(-E_s /k_B T)$, where $E_s$ is the energy of the state and $k_B T$ is the typical thermal energy; for our purposes we can choose units so that $k_B T = 1$.  Thus Eq (\ref{maxent}) defines an energy function or Hamiltonian on the space of the birds' velocities, and this can be written as
\be
{\cal H}(\{\vec{\mathbf{v}}_{\rm i}\})
=\frac{J}{4V^2} \sum_{{\rm ij}=1}^N n_{\rm ij} 
|\vec{\mathbf{v}}_{\rm i}-\vec{\mathbf{v}}_{\rm j}|^2 
+ \frac{g}{2 V^2}\sum_{{\rm i}=1}^N \left( v_{\rm i} - V\right)^2 ,
\label{hamiltonian}
\ee
where we have eliminated the parameter $\mu$ in favor of the mean speed $V$, which is now fixed to its experimental value $\langle V\rangle_{\rm exp}$, and we have set the arbitrary scale $v_0 = V$.

The first term in this Hamiltonian describes the tendency of the individual velocities to adjust both direction and modulus to their neighbors, while the second term forces the speed to have, on average, the  value $V$. From this perspective, we can interpret $J$ as the stiffness of an effective ``spring'' that ties each bird's velocity to that of its neighbors, and  $g$ as the stiffness of a competing spring that ties each speed to the desired mean.  Larger $J$ means a tighter connection to the neighbors, and larger $g$ means a tighter individual control over speed.

There are interesting limiting cases that give us a sense for what this model predicts.  If the parameter $g$ is very large, then the speed of individual birds hardly fluctuates at all.  In this limit, we can rewrite the Hamiltonian  as
\begin{equation}
{\cal H}(\{\vec{\mathbf{v}}_{\rm i}\})
\approx {\cal H}_{\rm dir}(\{\vec{\mathbf{s}}_{\rm i}\})
= -\frac{J}{2} \sum_{{\rm ij}=1}^N n_{\rm ij} \vec{\mathbf{s}}_{\rm i} \cdot \vec{\mathbf{s}}_{\rm j} ,
\label{dir_intuit}
\end{equation}
where $\vec{\mathbf{s}}_{\rm i}$ is the unit vector pointing in the direction that bird $\rm i$ is flying.  This Hamiltonian describes the tendency of the directions of individual birds to align with their neighbors, and is exactly the model in Ref \cite{bialek+al_12}.  

If there are nonzero but small fluctuations in speed, then we can write $v_{\rm i} = V (1+\epsilon_{\rm i})$, and expand in powers of $\epsilon$.  The result (Appendix \ref{app:solving}) is that
\begin{equation}
{\cal H}(\{\vec{\mathbf{v}}_{\rm i}\})
\approx  {\cal H}_{\rm dir}(\{\vec{\mathbf{s}}_{\rm i}\}) + {\cal H}_{\rm sp}(\{\epsilon_{\rm i}\}) ,
\label{sep1}
\end{equation}
where the ``speed Hamiltonian'' 
\begin{equation}
{\cal H}_{\rm sp} (\{ \epsilon_{\rm i}\}) = \frac{V^2}{2v_0^2}\sum_{\rm i,j =1}^N \left( g \delta_{\rm ij} + J N_{\rm ij}\right) \epsilon_{\rm i}\epsilon_{\rm j}  ,  \label{Hsp_a} 
\end{equation}
where the matrix $N_{\rm ij}$ has the form
\begin{equation}
N_{\rm ij} = - n_{\rm ij} + \delta_{\rm ij} \sum_{{\rm k}=1}^N n_{\rm ik} .
\end{equation}
Thus our full model breaks into two pieces, one of which describes fluctuations in flight direction, and one of which describes the fluctuations in speed.  Importantly, the strength of the ``springs'' that tie the speed of each bird to that of its neighbors is determined by the same parameter $J$ which enters the description of directional fluctuations in Eq (\ref{dir_intuit}).  Thus, we have a unified model for how birds adjust their vector velocities to those of their neighbors, rather than separate models (with separate parameters) for the adjustment of direction and speed.

To get a sense for the structure of $ {\cal H}_{\rm sp}$ it is useful to imagine a continuum limit, in which the variations in speed from bird to bird are so smooth that we can picture the speed fluctuations as a continuous function of position in the flock, $\epsilon (\vec{\mathbf{x}} )$.  In this limit (Appendix \ref{app:goldstone}), we have
\begin{equation}
{\cal H}_{\rm sp} \approx {\rho\over 2} \int d^3 x \left[ Jn_c r_c^2 (\nabla \epsilon )^2 + g \epsilon^2(\vec{\mathbf{x}} ) \right] ,
\end{equation}
where $r_c$ is the typical distance to a neighboring bird, and $\rho$ is the density of the flock.  This model predicts that the fluctuations will behave as
\begin{equation}
\langle \epsilon(\vec{\mathbf{x}} ) \epsilon(\vec{\mathbf{x}}' )\rangle \propto \exp\left( - |\vec{\mathbf{x}} - \vec{\mathbf{x}}'|/\xi_{\rm bulk}\right) ,
\label{exp_decay}
\end{equation}
where correlation length
\begin{equation}
\xi_{\rm bulk} \sim r_c \sqrt{{{J n_c}\over g}} 
\label{xi1}
\end{equation}
determines the distance over which the fluctuations in speed will be correlated;  the subscript reminds us that we are treating the flock as a bulk material, with no boundaries.  In this simple picture,  there is a critical point at $g=0$ where the correlation length $\xi_{\rm bulk}$ becomes infinite.

Thus the parameter $J$ determines the propagation of directional order through the flock, and to describe the speed fluctuations we have only one extra parameter $g$.  The value of $g$ is set by requiring that our model match the observed variance in speed across the birds in the flock, as in Fig \ref{fig:parameters}a.  But $J$ and $g$ also compete with one another to determine the distance over which speed fluctuations will be correlated, Eq (\ref{xi1}).  Importantly, we are not free to adjust this correlation length by some fitting procedure:  either the model gets it right, or it doesn't.

\section{Scale--free correlations}

\begin{figure*}[t] 
 \centering
  \includegraphics[scale=0.18]{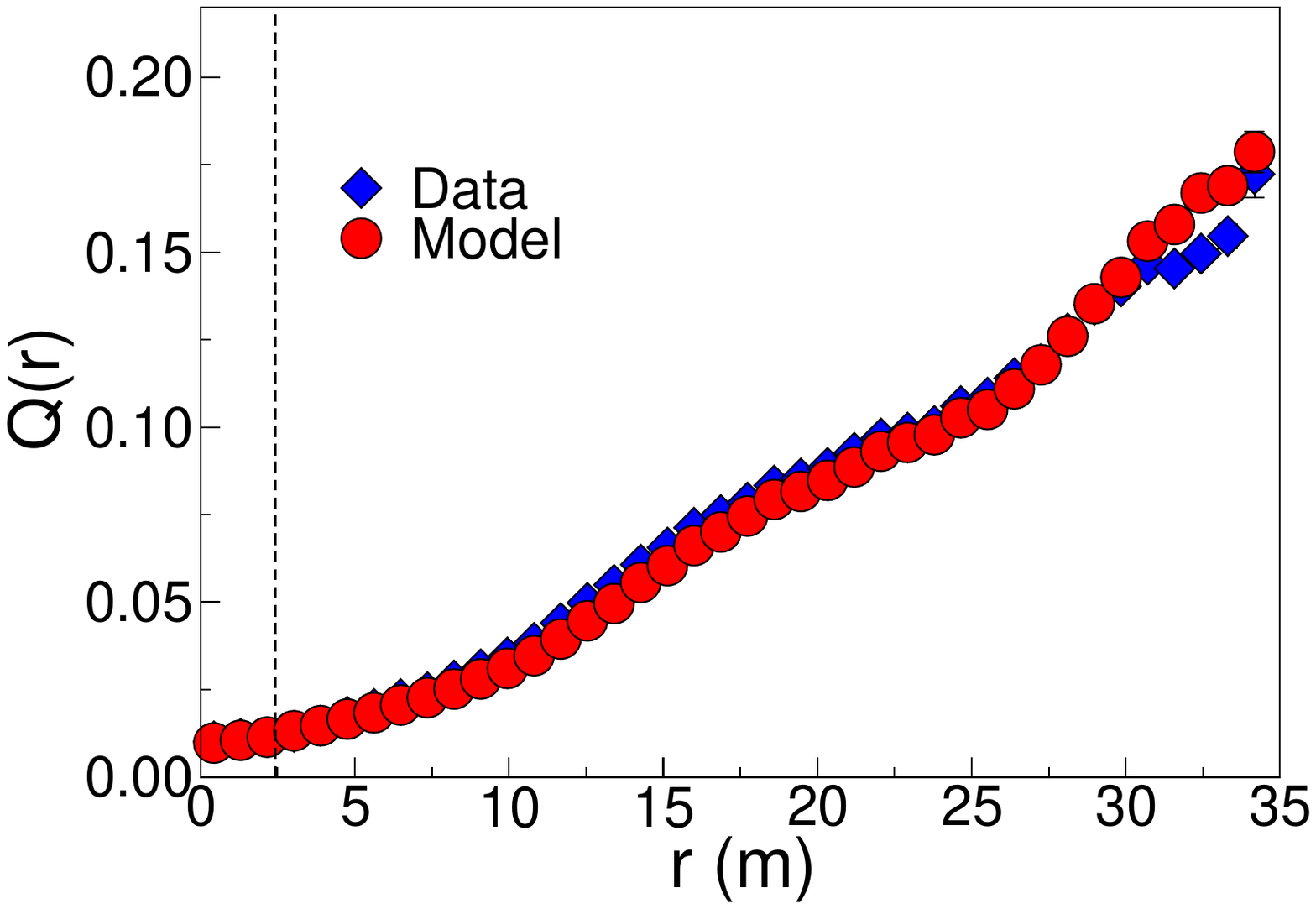}\quad
  \includegraphics[scale=0.18]{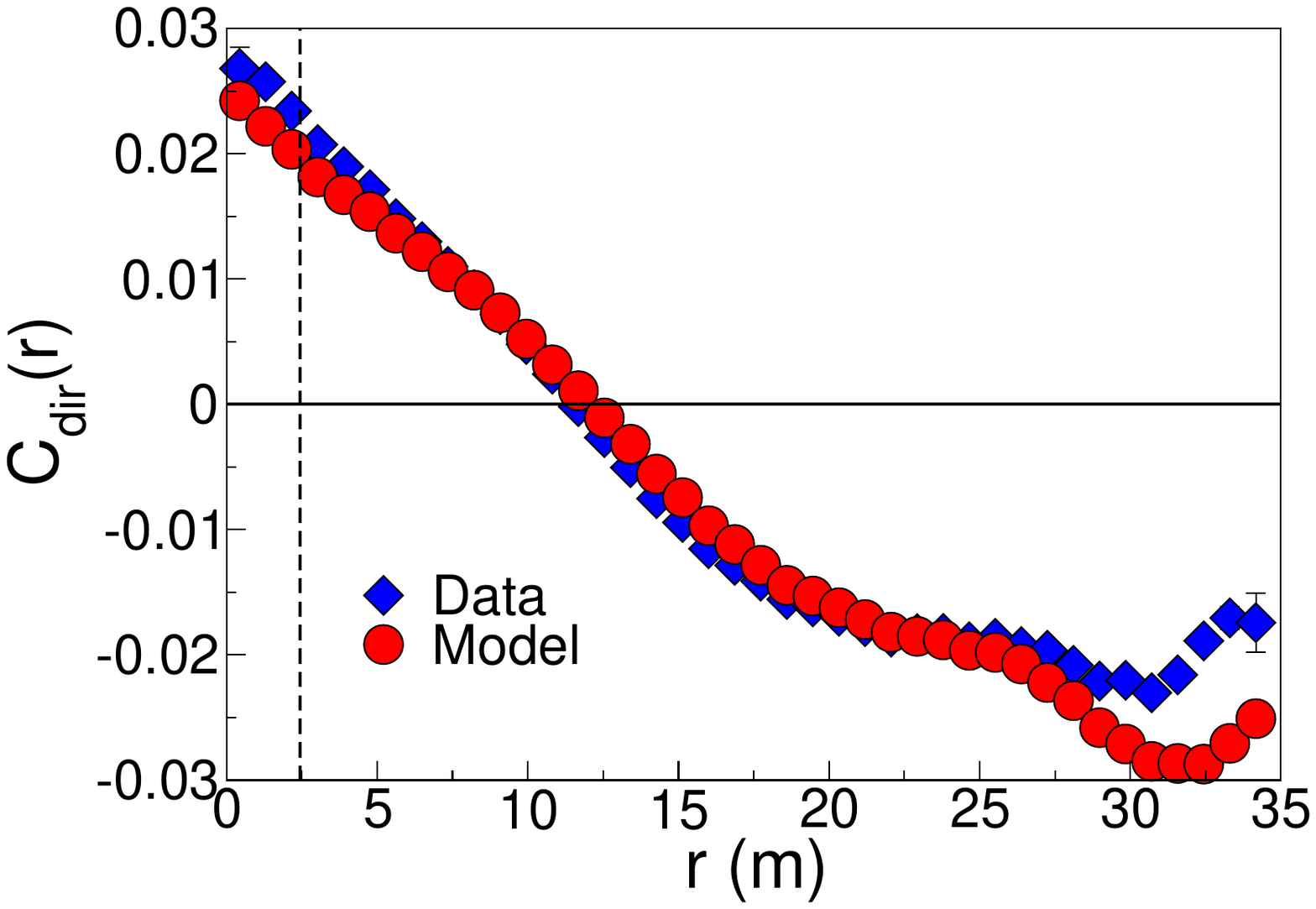}\quad
  \includegraphics[scale=0.18]{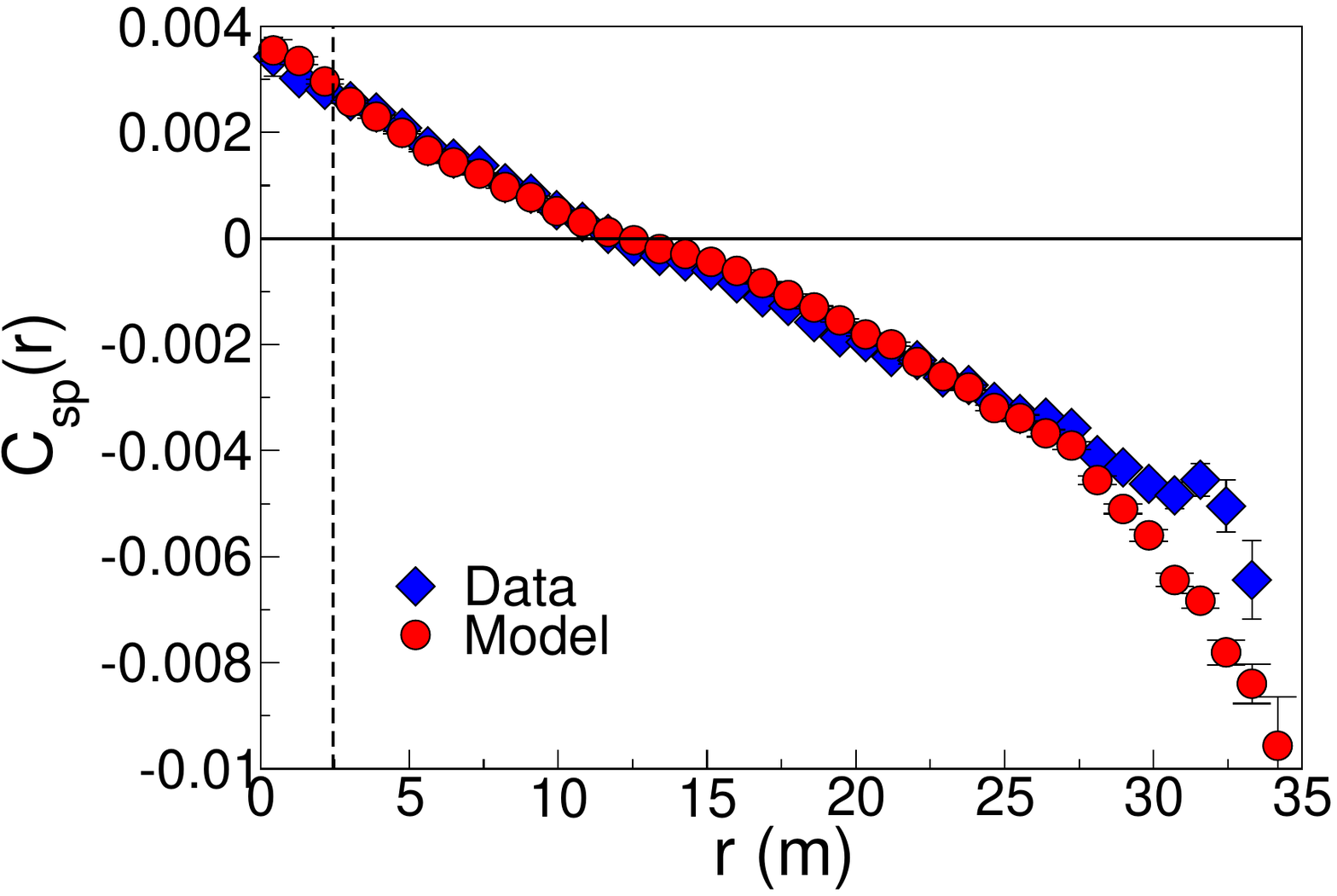}\\
   \quad\mbox{   (a)}\hspace{5.5 cm}\mbox{(b)}\hspace{5.5 cm}\mbox{(c)}
\caption{Correlation functions predicted by the maximum entropy model (red circles) vs. experiments (blue diamonds).  (a) Similarity of velocities as a function of distance, defined in Eq (\ref{Qdef}).  Dashed line indicates the size of the neighborhood defined by $n_c$ birds, within which we match the average $Q$ exactly, by construction.   (b) Correlations between fluctuations in flight direction as a function of distance, defined in Eq (\ref{Cdir}).   (c) Correlations between fluctuations in speed as a function of distance, defined in Eq (\ref{Csp}).   All panels refer to the same flock and snapshot as in Fig \ref{fig:parameters}; theoretical predictions from Appendix \ref{fixed}.    \label{fig:correlations}}
\end{figure*}

Once the parameters $J$, $g$ and $n_c$ are determined (Fig \ref{fig:parameters}), Eq (\ref{maxent}) provides  a model for the joint distribution of velocities for all the birds in the flock; everything that we compute from this distribution is a parameter--free prediction.   We start by measuring the similarity of the vector velocities among birds that are not just nearest neighbors, but are separated by greater distances. By analogy with Eq (\ref{Qint}),  we can define
\begin{equation}
Q(r) = {1\over{V^2}} \langle |\vec{\mathbf{v}}_{\rm i}-\vec{\mathbf{v}}_{\rm j}|^2 \rangle_{r_{\rm ij} = r} ,
\label{Qdef}
\end{equation}
where the average is over all pairs of birds in the flock separated by a distance $r_{\rm ij} = r$.    We see in Fig \ref{fig:correlations}a that the predicted $Q(r)$ matches the data very closely, out to distances comparable to the overall size of the flock, more than ten times farther than the nearest neighbors.

We next decompose the relationships among velocities into contributions from direction and speed.  If we take all the units vectors $\vec{\mathbf{s}}_{\rm i}$ and average, we obtain the overall polarization of the flock,
\begin{equation}
\vec{\mathbf{P}} = {1\over N}\sum_{{\rm i}=1}^N \vec{\mathbf{s}}_{\rm i} ,
\label{Pdef}
\end{equation}
and we can characterize the fluctuations around this overall direction by a correlation function
\begin{equation}
C_{\rm dir}(r) = {\bigg\langle} \left( \vec{\mathbf{s}}_{\rm i} - \vec{\mathbf{P}}\right) \cdot \left(\vec{\mathbf{s}}_{\rm j} -\vec{\mathbf{P}}\right) {\bigg\rangle}_{r_{\rm ij} = r} .
\label{Cdir}
\end{equation}
In Fig \ref{fig:correlations}b we compare the data with the predictions of the model, and again find that the agreement is very good, on all scales.

By analogy with Eq (\ref{Cdir}), we can define correlations among the fluctuations in speed,
\begin{equation}
C_{\rm sp}(r) = {\bigg\langle} \left( v_{\rm i} - V \right) \cdot \left( v_{\rm j} -V \right) {\bigg\rangle}_{r_{\rm ij} = r} .
\label{Csp}
\end{equation}
Fig \ref{fig:correlations}c shows that the observed correlations are in   agreement with the predictions of the model, again over the full range of distances.  Thus, we have succeeded in constructing  a model based on local interactions that generates  correlations in speed fluctuations over long distances, matching the data quantitatively.

\begin{figure}[b] 
\centering
\includegraphics[scale=0.22]{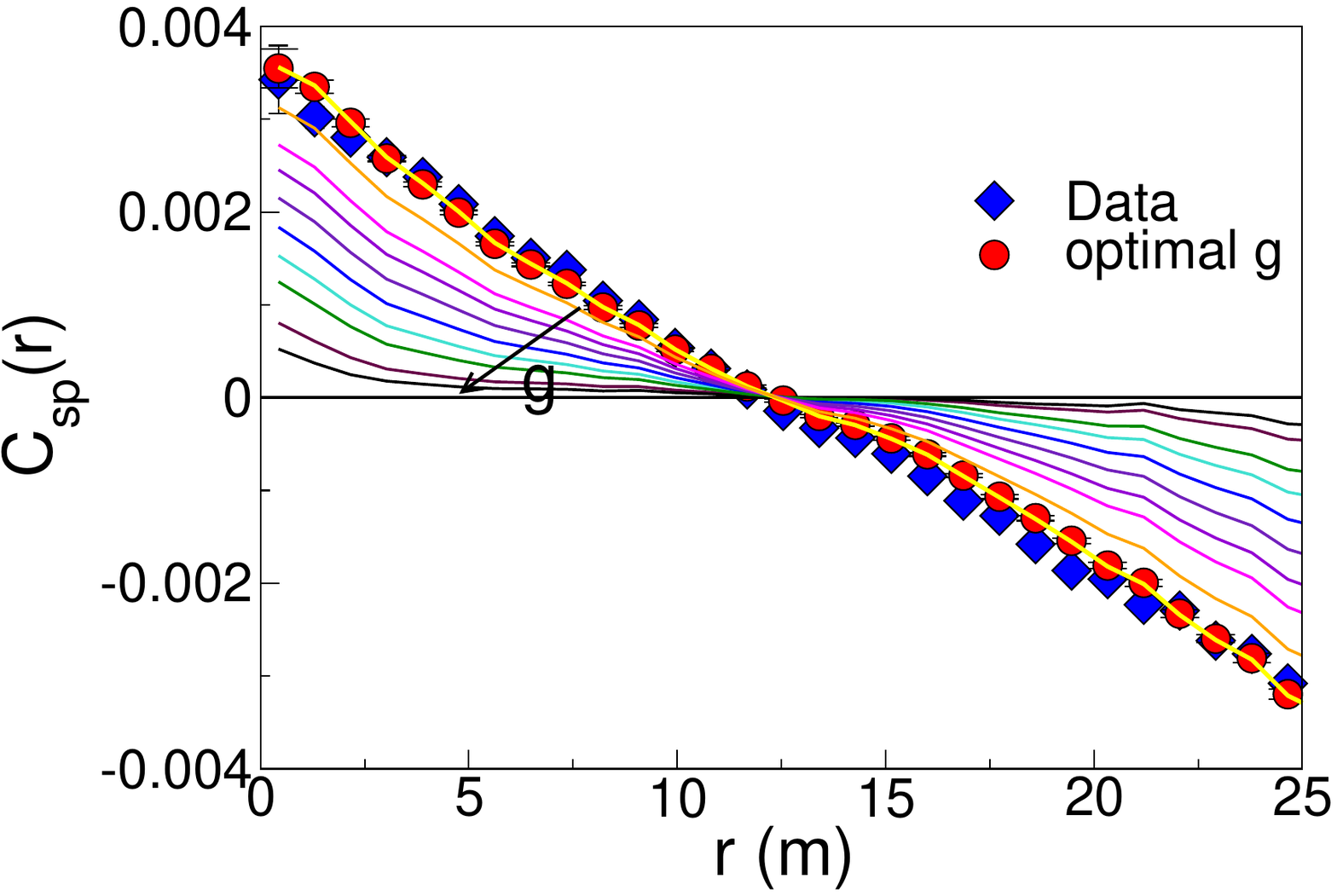}\\ \quad\quad \mbox{(a)}\\
 \quad \quad\quad\includegraphics[scale=0.23]{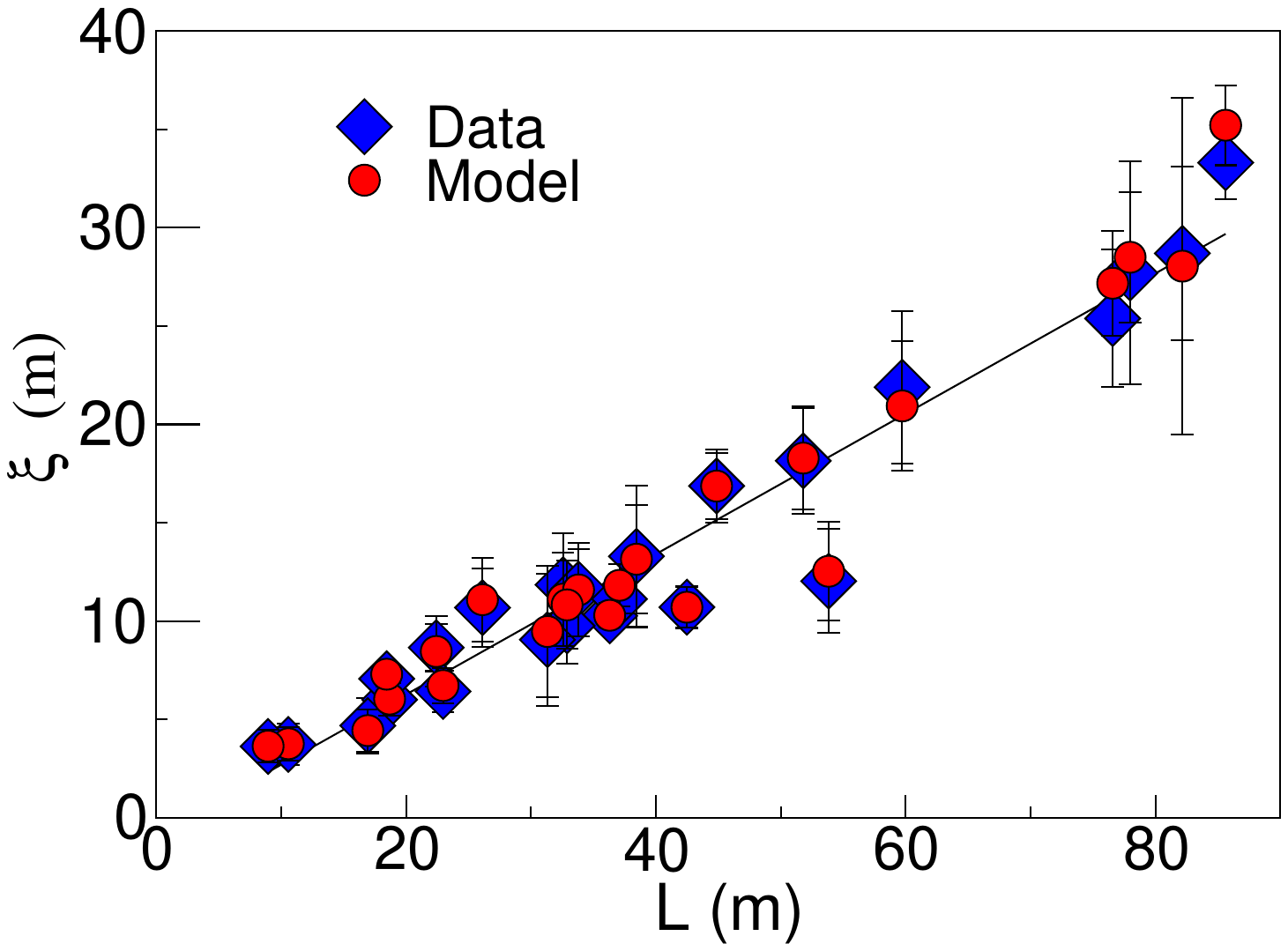} \\ \quad\quad\mbox{(b)}
  \caption{ (a) Correlation function of the speed fluctuations, for different values of the control parameter $g$ (increasing in the direction of the arrow). (b) Correlation length, defined as the point where the correlation function crosses zero \cite{cavagna+al_10}, in flocks of different sizes, for the experimental data (blue diamonds) and for the model (red circles).    \label{fig:multi-g}}
\end{figure}

The discussion in Section III suggests that  long ranged correlations are associated with the approach to a critical point at $g=0$.  To see if this intuition is correct, we show in Fig \ref{fig:multi-g}a what happens to the predicted $C_{\rm sp}(r)$ as we change the value of $g$.  Large values of $g$ correspond to small variances in speed, and to correlation functions that decay very rapidly with distance.  As $g$ becomes smaller, both the speed variance and the correlation length increase, until, for sufficiently small $g$, there really isn't a characteristic scale to the decay of the correlations, and $C_{\rm sp}(r)$ is almost a straight line.  This is the shape of the correlation function we observe, and the success of the theory is that the value of $g$ that matches the observed speed variance is in this regime. 

We can quantify the approach to criticality by the dimensionless ratio $g/(Jn_c)$ that enters Eq (\ref{xi1}).  From Fig \ref{fig:parameters}, we see that $g/(Jn_c)\sim 10^{-3}$, and this is typical. This suggests that real flocks are very close to criticality, and that this is why we observe scale--free speed correlations. Note that  $g$ cannot be exactly zero, otherwise the variance in speed would be infinite; a non--zero (even if small) value of $g$ is  necessary to fix the flock's speed.

To be more precise about the approach to criticality, we  need to take into account the finite size of the flocks.  Equations (\ref{exp_decay}) and (\ref{xi1})  hold only for an infinite system;  for a finite system, the range of the correlation cannot increase indefinitely, since it is limited by the system size. As $g$ is lowered, the behavior of the correlations is   influenced more and more by these finite size effects: the exponential decay in Eq (\ref{exp_decay})  is modified, and the typical distance over which correlations extend  is no longer described by $\xi_{\rm bulk}$. A more faithful estimate of the correlation length $\xi$ is  given instead by the zero of the correlation function \cite{cavagna+al_10}, and the theoretical prediction depends in a non--trivial way on $g$ and the system size $L$. For small enough values of $g$, however,  the system is effectively critical and scale--free;  we should see $\xi \propto L$. In Fig \ref{fig:multi-g}a we show that decreasing $g$ below the level required to match the speed variance of the real flock has essentially no effect, and curves with all smaller values of $g$ ``pile up'' as shown in yellow.  Repeating the analysis on flocks of different sizes (Fig \ref{fig:multi-g}b),  we see that the correlation length does scale with size,  and this pattern is captured perfectly by our maximum entropy models. 

We conclude that flocks do in fact exhibit critical behavior, being close enough to the critical point  to achieve  maximum speed correlation length while still maintaining a well defined cruising speed and limited speed fluctuations. These conclusions are rather robust, and also hold when considering more general maximum entropy models where speed and flight directions are regulated by different interaction parameters (see Appendix \ref{app:decoupling}).

\section{Dynamical model}

The fact that maximum entropy models are equivalent to the Boltzmann distribution for a system in thermal equilibrium suggests a natural dynamical model, in which the various degrees of freedom in the system execute Brownian motion on the energy landscape.  We can describe such dynamics with a Langevin equation,
\be
\gamma \frac{d\vec{\mathbf{v}}_{\rm i}(t)}{dt}
=
-\nabla_{\rm i} {\cal H}(\{\vec{\mathbf{v}}_{\rm j}\}) +\vec{\mathbf{\eta}}_{\rm i} (t) ,
\label{dynamics}
\ee
where $\nabla_{\rm i}$ indicates the derivatives with respect to the components of the velocity $\vec{\mathbf{v}}_{\rm i}$, $\gamma$ is a constant to set the time scale of the dynamics, and the Langevin force $\vec{\mathbf{\eta}}_{\rm i} (t)$ is a random, white noise function of time.  These dynamics are guaranteed, if we assume that the positions of the birds are fixed, to generate velocities that are drawn from the probability distribution in Eq (\ref{maxent}).  But to give a more realistic model we should add to Eq (\ref{dynamics}) forces that depend on the positions of the birds \cite{gregoire+chate_04,camperi+al_12,oliver}, so as to fix the overall density of the flock (see Appendix \ref{app:dynamics}), and the velocities should drive the birds' positions,
\be
\frac{d\vec{\mathbf{x}}_{\rm i}}{dt}=\vec{\mathbf{v}}_{\rm i} .
\label{kinetic}
\ee
Equations (\ref{dynamics}) and (\ref{kinetic}) define a  ``self--propelled particle'' model of interacting birds, and is similar to the Vicsek model,  so often used to describe flocking particles \cite{vicsek+al_95,gregoire+al_03}. In contrast to that model, and to most of flocking models in the literature, the speed of the individual particles is not fixed, but regulated by the control parameter $g$.

\begin{figure}[tb] 
 \centering
\includegraphics[scale=0.22]{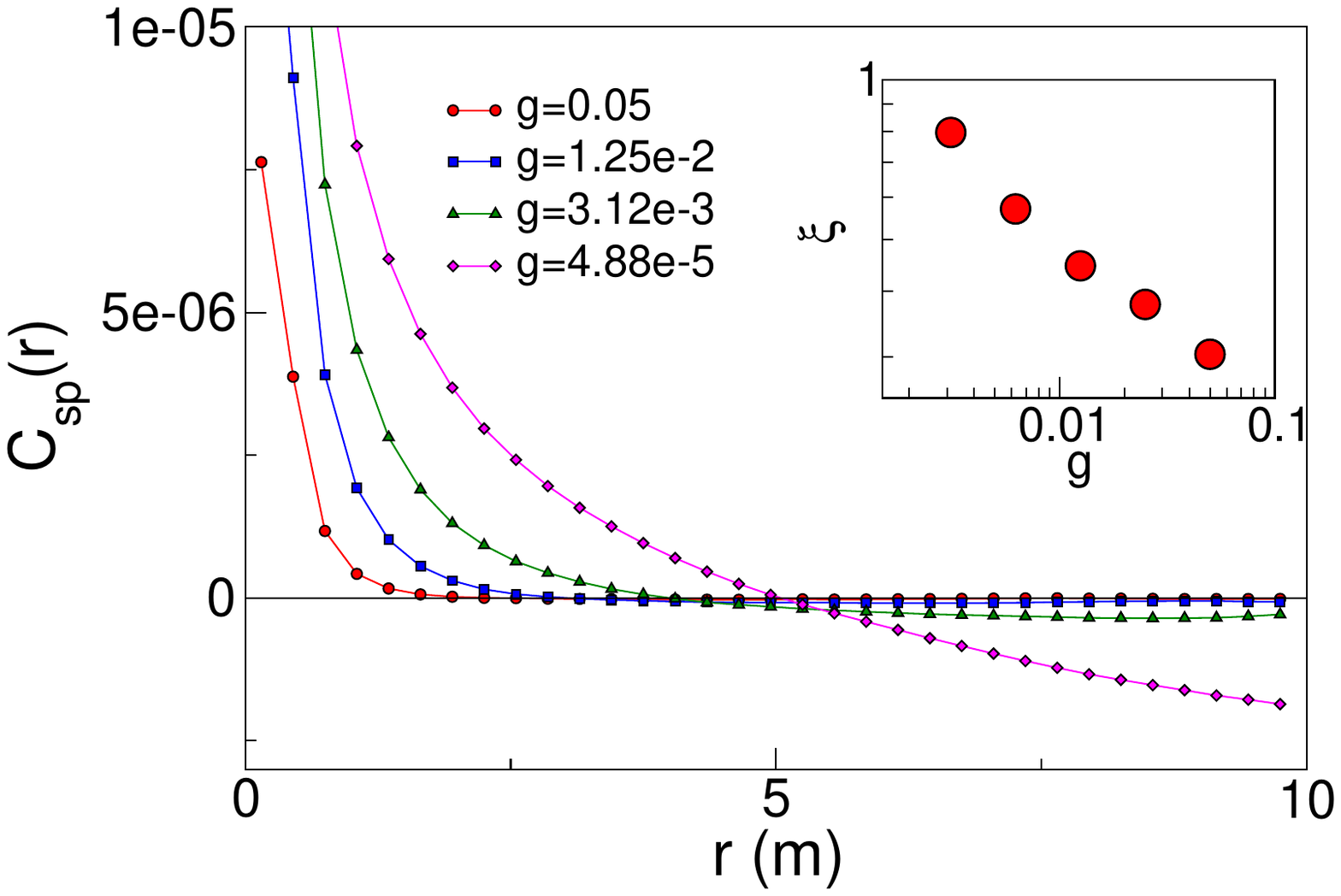}\\ \quad\quad\mbox{(a)} \\
\quad  \includegraphics[scale=0.22]{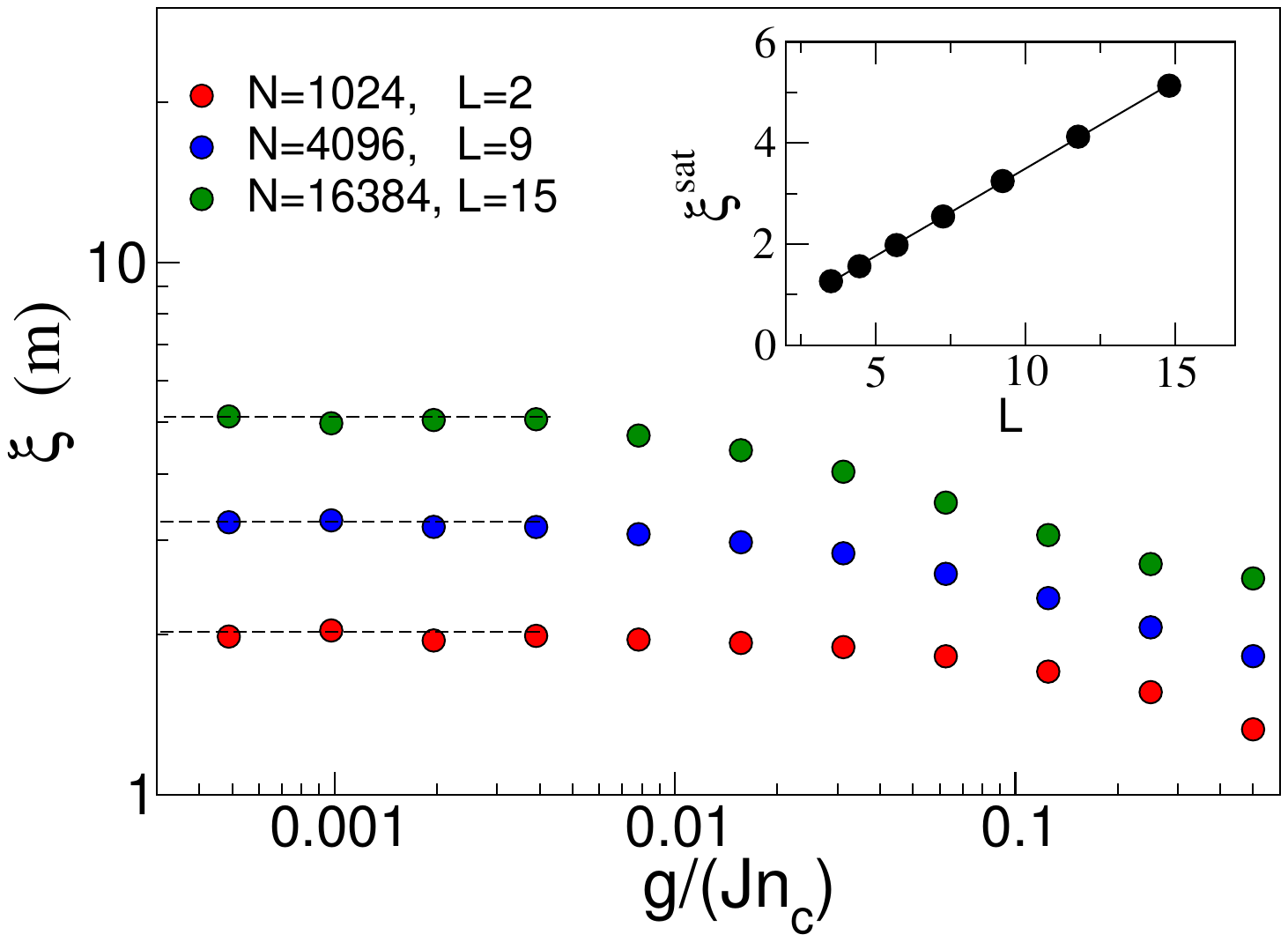} \\ \quad\quad \mbox{(b)}
\caption{Simulations of a dynamical model (see Appendix for details). (a) Correlation function of the speed fluctuations at different values of $g$, in a flock of $N=16384$ birds. Inset: Correlation length, measured from the exponential decay of the correlation functions at small $r$,  as a function of $g/(Jn_c)$. (b) For smaller $g$, correlation lengths are measured from the zero crossing of the correlation function.  For $g/(Jn_c)\ll1$,  $\xi$  approaches a maximum value that depends on the size of the system. Inset: low-$g$ maximum of $\xi$, as a function of the system size; the linear dependence of $\xi$ on $L$ is typical of scale--free behavior.     \label{fig:dynamics}}
\end{figure}

Simulations of the dynamical model defined by Eqs (\ref{dynamics}, \ref{kinetic}) are shown in Fig \ref{fig:dynamics}.  As expected from the analysis of the (static) maximum entropy model, the fluctuations in speed have a correlation length that grows as $g$ is reduced.  If $g$ is not too small, we see correlations that decay exponentially [Eq (\ref{exp_decay})], and the correlation length varies with $g/(Jn_c)$ as expected.  When $g$ is lowered even further, the exponential decay is modified by finite size corrections, and the correlation length---now computed as the zero--crossing point of the correlation function---keeps decreasing until  a maximal,  size dependent saturation value is reached.  In this regime, the  correlations extend over a distance   determined by the system size, and $\xi$ in fact grows linearly with $L$  corresponding to scale--free behavior (Fig \ref{fig:dynamics}b, inset). This scenario confirms that the mechanism identified in the previous section  produces scale--free correlations in the speed even when the full dynamical behavior of the flock is taken into account.

\section{Conclusions}

The understanding of collective behavior in matter at thermal equilibrium provide a touchstone for thinking about emergent  phenomena in complex, biological systems.  Flocking seems like an especially attractive example, in which the alignment of birds in a flock reminds us of the alignment of spins in a magnet or molecules in a liquid crystal.  But birds are vastly more complex than spins, and this might be nothing more than a  metaphor.  The goal of this paper and its companion \cite{bialek+al_12} has been to show that we can go beyond metaphor, that there is a statistical mechanics description of flocks which makes quantitative, parameter--free predictions that are in detailed agreement with the data.

One dramatic collective phenomenon that can emerge in statistical mechanics is the existence of a critical point.  At such points, distant elements of a system become correlated with one another, far beyond the range of local interactions among the individual elements.  At generic parameter values, correlations are expected to decay on some characteristic spatial scale $\xi$, so that a very large system is composed of many nearly independent pieces of volume $\xi^3$; often, $\xi$ is not much larger than the range of the interactions themselves.  But at a critical point, the correlation length $\xi$ becomes (formally) infinitely large, and the scale over which correlations extend becomes comparable to the linear size $L$ of the entire system; rather than having many independent pieces, the system acts (almost) as one.

The idea that biological systems might tune themselves to critical points is not new \cite{bak}, but has languished for lack of detailed comparison with experiment.    The emergence of new and more extensive data, as well as new ideas about how to connect theory and experiment, has led to a re--examination of criticality in a wide variety of biological systems \cite{mora+bialek_12}.  In this context, the observation of long ranged, or scale--free correlations in the velocities of starlings in a flock \cite{cavagna+al_10} is very suggestive.  Our results here show that these correlations are not just analogous to the correlations at a critical point:  we have a very accurate description of the entire distribution of speed and direction fluctuations in the flock, this description is mathematically equivalent to a statistical mechanics model of a magnet, and the observed scale--free correlations are predicted correctly because the parameters of this model  are in the critical regime.

Our approach is not a ``fit'' to the observed scale--free correlations in the flock.   Instead we take from the data a measurement of local correlations, and the variance of individual birds' speeds relative to the average over the flock, and build the least structured model that is consistent with these two measurements.    Thus, rather than thinking of criticality as occurring in the neighborhood of a special point in the space of model parameters, we can think of it as a statement about the behavior of the flock itself.  In particular, as emphasized in Fig \ref{fig:multi-g}, even a factor of two change in the variance of the speeds would predict correlations that decay much more rapidly with distance, inconsistent with what we see in real flocks.  

Biologically, birds may vary their speeds either for individual reasons \cite{rainer+al_01}, or to follow their neighbors, paralleling the competing forces captured in the model.  In this language, the critical point is the place where social forces overwhelm individual preferences.  More broadly, the critical regime is one in which is individuals achieve maximal coherence with their neighbors while still keeping some control over their speeds.  

Why do flocks organize themselves to be critical?  Historically, there has been much more speculation about the advantages of criticality for biological systems than there has been direct evidence, so we do not want to add too much here.  We note, however, that in the statistical mechanics framework, the long ranged correlations at criticality are mathematically equivalent to the statement that information can propagate over similarly long distances.  Away from criticality, a signal visible only to one bird on the border of the flock can influence just a handful of near neighbors; at criticality, the same signal can spread to influence the behavior of the entire flock. Such susceptibility seems advantageous in terms of anti--predatory strategies, but it would be attractive to have more direct measurements of the propagating signal \cite{attanasi+al_13}.  The critical point is a place where many quantities are extremal; it remains to be seen which of these is most meaningful to the birds.

\begin{acknowledgments}
We thank G Tka\v{c}ik and G Parisi for many helpful discussions. Work in Princeton was supported in part by National Science Foundation Grants  PHY--0957573 and CCF--0939370, and by the WM Keck Foundation; work in Rome was supported in part by grants IIT--Seed Artswarm, ERC--StG n. 257126,  US--AFOSR grant FA95501010250 (through the University of Maryland); work in Paris was supported by grant ERC--StG  n. 306312. Our collaboration was facilitated by the Initiative for the Theoretical Sciences at the Graduate Center of the City University of New York. 
\end{acknowledgments}

\appendix

\section{Data}
\label{app:data}

The data that we analyze here were obtained from observations on large flocks of starlings, {\em Sturnus vulgaris}, in the field. Using stereometric photography and innovative computer vision techniques \cite{cavagna+al_08a,cavagna+al_08b} the individual 3D coordinates  and velocities were measured in cohesive groups of up to 4268 individuals \cite{ballerini+al_08a,cavagna+al_10}. As summarized in Table \ref{table}, we have data from 21 distinct flocking events, with sizes ranging from 122 to 4268 individuals and linear extensions from $9.1$ to $85.7\,{\rm  m}$. Each event consists of up to 40  consecutive 3D configurations (individual positions and velocities), at time intervals of $1/10\,{\rm s}$.  All events correspond to strongly ordered flocks, with polarization [from Eq (\ref{Pdef})] between $|\vec{\bf P}| = 0.844$ and $|\vec{\bf P}| = 0.992$. The border of each flock at each instant of time has been computed using the $\alpha$--shape algorithm \cite{edelsbrunner_94}, as explained in detail in \cite{cavagna+al_08b}.

\begin{table}[h]
\vspace{0.5 cm}
\begin{tabular}{@{\vrule height 10.5pt depth 4pt  width0pt}|c|c|c|c|c|c|}
\hline
 Event & $N$ & $P$ & $\langle V\rangle_{\rm exp}$ (${\rm m}/{\rm s}$) & $L$ ($\rm m$) &  $Q_{\rm int}$ \\
\hline
17--06 & 552  & 0.935 &  9.96  & 51.8 & 1.29e-01\\
\hline
21--06 & 717  & 0.973 & 12.06  & 32.1 & 1.22e-02 \\
\hline
25--08 & 1571 & 0.962 & 12.47  & 59.8 & 2.63e-02 \\
\hline
25--10 & 1047 & 0.991 & 12.57  & 33.5 & 8.36e-03 \\
\hline
25--11 & 1176 & 0.959 & 10.07  & 43.3 & 6.27e-02 \\
\hline
28--10 & 1246 & 0.982 & 11.22  & 36.5 & 6.43e-03 \\
\hline
29--03 & 440  & 0.963 & 10.75  & 37.1 & 1.43e-02 \\
\hline
31--01 & 2126 & 0.844 &  8.13  & 76.8 & 5.50e-02\\
\hline
32--06 & 809  & 0.981 &  9.99  & 22.2 & 1.52e-02\\
\hline
42--03 & 431  & 0.979 & 10.68  & 29.9 & 1.62e-02\\
\hline
49--05 & 797  & 0.995 & 14.02  & 19.2 & 6.49e-03\\
\hline
54--08 & 4268 & 0.966 & 19.17  & 78.7 & 4.29e-02\\
\hline
57--03 & 3242 & 0.978 & 14.38  & 85.7 & 1.53e-02\\
\hline
58--06 & 442  & 0.984 & 10.13  & 23.1 & 1.34e-02 \\
\hline
58--07 & 554  & 0.977 & 10.81 & 19.1 & 1.35e-02\\
\hline
63--05 & 890  & 0.978 & 10.24 & 52.9 & 1.86e-02\\
\hline
69--09 & 239  & 0.985 & 11.97 & 17.1 & 2.68e-02\\
\hline
69--10 & 1129 & 0.987 & 12.04 & 47.3 & 2.35e-02\\
\hline
69--19 & 803  & 0.975 & 14.16 & 26.4 & 3.65e-02 \\
\hline
72--02 & 122  & 0.992 & 13.24 & 10.6 & 1.12e-02 \\
\hline
77--07 & 186  & 0.978 &  9.50 & 9.1 & 4.27e-02 \\
\hline \end{tabular}
\caption{Summary of experimental data.   Flocking events are labelled according to experimental session number and to the position within the session to which they belong. The number of birds $N$ is the number of individuals for which we obtained a 3D reconstruction of positions in space. The polarization $P$  is  the global degree of alignment, as defined in the text. The linear size $L$ of the flock is defined as the maximum distance between two birds belonging to the flock. The speed $\langle V \rangle_{\rm exp}$ is the average of the individual speeds over all the individuals in the flock, and $Q_{\rm int}$ is as defined in Eq (\ref{Qint}). All values are averaged over several snapshots during the flocking event.  \label{table} }
\end{table}

\section{The maximum entropy approach}
\label{maxentapp}

The concept of entropy has its roots in thermodynamics, roughly 150 years ago.  The idea that we can use maximum entropy as a strategy to construct simplified models outside of equilibrium thermodynamics is now more than 50 years old \cite{jaynes_57}.  Here, so that our discussion is self--contained, we review this general strategy.  See also Ref \cite{bialek+al_12}, and Appendix A.7 of Ref \cite{bialek_12}.

We assume that the state of the system can be described by a set of variables that we shall call ${\bf v}\equiv \{ \vec{\mathbf{v}}_1,\vec{\mathbf{v}}_2\cdots, \, \vec{\mathbf{v}}_N\}$, by analogy with the velocities of birds in a flock.   Although we can measure, for example, the velocity of every bird in a flock, we typically can't collect enough data to make reliable estimates of very complicated quantitates.  As an example, with $N$ variables describing the state of the system, we need more than $N$ independent measurements to be sure that the covariance matrix of these variable is not artificially singular.  What does seem reasonable is to assume that there is a much smaller set of observables,  $\{ O_\mu({\bf v})\}$ with $\mu=1,\, 2,\, \cdots ,\, K$, that we can extract from the system, and that we have enough data to make reliable statements about the average values of these obervables, $\{\langle O_\mu({\bf v})\rangle_{\rm exp}\}$.  

Our task is to build a probability distribution $P({\bf v})$ such that we reproduce, exactly, the expectation values of the $K$ observables, that is
\begin{equation}
\langle O_\mu({\bf v})\rangle_P \equiv  \sum_{\bf v} P({\bf v}) O_\mu({\bf v}) = \langle O_\mu({\bf v})\rangle_{\rm exp} ,
\label{fixedOs}
\end{equation}
for all $\mu=1,\, 2,\, \cdots ,\, K$; it is useful to phrase the normalization of the distribution as a similar constraint,  the statement that the average of the ``function'' $O_0 ({\bf v}) = 1$ must equal the ``experimental'' value of $1$.

The problem is that that there are infinitely many distributions that can satisfy the constraints in Eq (\ref{fixedOs}).  Out of all these distributions, we want to find the one that has as little structure as possible, so that we can derive the minimal consequences of the experimental observations on $\{\langle O_\mu({\bf v})\rangle_{\rm exp}\}$.  Asking for a probability distribution $P({\bf v})$ that has as little structure as possible is equivalent to asking that the variables $\bf v$ that we draw out of this distribution be as random as possible. Shannon proved that the only measure of (lack of) structure or randomness that is consistent with several simple constraints is the entropy of the distribution \cite{shannon_48,cover+thomas_91},
\begin{equation}
S\left [P\right ]=-\sum_{\bf v} P({\bf v})\ln P({\bf v}) .
\label{entropy}
\end{equation}
Thus, we are looking for the distribution $P({\bf v})$ that maximizes the entropy in Eq (\ref{entropy}) while obeying the experimental constraints from Eq (\ref{fixedOs}).  Such   constrained optimization problems can be solved using the method of the Lagrange multipliers \cite{bender+orszag}:  we introduce a generalized entropy function,
\begin{equation}
{\cal S}\left[ P;\{\lambda_\nu\}\right]=S\left [P\right ]-\sum_{\mu =0}^K  \lambda_{\mu} \left [ \langle O_{\mu}({\bf v})\rangle_P - \langle O_{\mu}({\bf v})\rangle_{\rm exp} \right ] ,
\label{gen_entropy}
\end{equation}
where a multiplier $\lambda_\mu$ appears for each constraint to be satisfied, and then we maximize $\cal S$ both with respect to the probability distribution $P({\bf v})$ and with respect to the parameters $\{\lambda_\mu\}$.  

Maximizing with ${\cal S}$ respect to $P({\bf v})$ gives
\be
P({\bf v}) = {1\over {{\cal Z}(\{\lambda_\nu\})}} \exp\left[ - \sum_{\mu = 1}^K \lambda_\mu O_\mu ({\bf v})\right] ,
\label{P_maxent}
\ee
where ${\cal Z}(\{\lambda_\nu\}) = \exp(-\lambda_0 - 1)$.  Since optimizing with respect to $\lambda_0$ will enforce normalization of the distribution, we can write, explicitly,
 \begin{equation}
{\cal Z}(\{\lambda_\nu\}) = \sum_{\bf v} \exp\left[ - \sum_{\mu = 1}^K \lambda_\mu O_\mu ({\bf v})\right] .
\label{ZA}
\end{equation}
Maximizing with respect to $\{\lambda_\nu\}$ gives us the set of $K$ simultaneous equations in Eq (\ref{fixedOs}), which we can now write more explicitly as
\be
 \langle O_\mu ({\bf v})\rangle_{\rm exp} = {1\over {{\cal Z}(\{\lambda_\nu\})}} \sum_{\bf v}  O_\mu ({\bf v})\exp\left[ - \sum_{\nu = 1}^K \lambda_\nu O_\nu ({\bf v})\right]  \ .
\label{multipliers}
\ee
We note that, in general, this is a very nonlinear set of equations for the parameters $\{\lambda_\nu\}$, and very hard to solve.  In the next section we exploit special features of the flock problem---in particular, the strong polarization of the flock---to simplify this problem so that we can make analytic progress.

Maximum entropy distributions are mathematically equivalent to the Boltzmann distribution in statistical physics.  We recall that if  a physical system in state $\bf v$ has energy $E({\bf v})$, then when it comes to equilibrium at temperature $T$ the probability that is in any particular state is given by
\begin{equation}
P_{\rm Boltz} ({\bf v}) = {1\over {Z}} \exp\left[ - {{E({\bf v})}\over {k_B T}}\right],
\end{equation}
where $k_B$ is Boltzmann's constant, and serves to convert between conventional units of temperature and energy.  Comparing with Eq (\ref{P_maxent}), we see that the maximum entropy distribution is equivalent to a Boltzmann distribution with an effective energy
\begin{equation}
{{E({\bf v})}\over {k_B T}} = \sum_{\mu = 1}^K \lambda_\mu O_\mu ({\bf v}) .
\label{app_energy}
\end{equation}
We note that this energy is the sum of several terms, one for each of the observables whose expectation value we fix based on experimental data.

It also is useful to note the connection of the maximum entropy approach to more conventional model building.  If we take the form of the probability distribution in Eq (\ref{P_maxent}) as given, then our problem is only to ``fit'' the parameters $\{\lambda_\nu\}$.  A standard method is maximum likelihood.  If we have $N_s$ independent samples of the system's state, ${\bf v}^{(1)},\  {\bf v}^{(2)}, \, \cdots ,\, {\bf v}^{(N_s)}$, then the probability that the model generates these data is given by
\begin{equation}
P_{\rm model} ({\rm data}) = \prod_{{\rm i}=1}^{N_s} P({\bf v}^{({\rm i})}) .
\end{equation}
Substituting from Eq (\ref{P_maxent}) we can make this more explicit,
\begin{widetext}
\begin{equation}
P_{\rm model} ({\rm data})= 
{1\over{{\cal Z}^{N_s}(\{\lambda_\nu\})}} \prod_{{\rm i}=1}^{N_s}
\exp\left[ - \sum_{\mu = 1}^K \lambda_\mu O_\mu ({\bf v}^{({\rm i})})\right] =  {1\over{{\cal Z}^{N_s}(\{\lambda_\nu\})}}
\exp\left[ -\sum_{\mu = 1}^K \lambda_\mu \sum_{{\rm i}=1}^{N_s}O_\mu ({\bf v}^{({\rm i})})\right] .
\end{equation}
Then we can form the normalized log probability,
\begin{eqnarray}
{1\over {N_s}}\ln P_{\rm model} ({\rm data}) &=& -\ln {\cal Z}(\{\lambda_\nu\}) 
- \sum_{\mu = 1}^K \lambda_\mu \left[ {1\over {N_s}}\sum_{{\rm i}=1}^{N_s}O_\mu ({\bf v}^{({\rm i})})\right] \label{logP1}\\
&=& -\ln {\cal Z}(\{\lambda_\nu\}) 
- \sum_{\mu = 1}^K \lambda_\mu  \langle O_\mu ({\bf v})\rangle_{\rm exp},
\end{eqnarray}
where in the last step we recognize the normalized sum over samples as the experimental expectation value.    Now if we want to maximize the probability, or likelihood, we should differentiate with respect to the parameters and set the result to zero:
\begin{equation}
{{\partial \ln P_{\rm model}({\rm data})}\over{\partial \lambda_\mu}} = 0  
\Rightarrow {{\partial \ln {\cal Z}(\{\lambda_\nu\})}\over{{\partial \lambda_\mu}} }
= -\langle O_\mu ({\bf v})\rangle_{\rm exp}.\label{maxlike1}
\end{equation}
But with the explicit expression for $\cal Z$ in Eq (\ref{ZA}), we can compute:
\begin{eqnarray}
{{\partial \ln {\cal Z}(\{\lambda_\nu\})}\over{{\partial \lambda_\mu}} }
&=& {1\over{ {\cal Z}(\{\lambda_\nu\})}} {{\partial  {\cal Z}(\{\lambda_\nu\})}\over{{\partial \lambda_\mu}} }=  {1\over{ {\cal Z}(\{\lambda_\nu\})}} {\partial\over{\partial\lambda_\mu}} 
\sum_{\bf v} \exp\left[ - \sum_{\nu = 1}^K \lambda_\nu O_\nu ({\bf v})\right] \\
&=& - {1\over{ {\cal Z}(\{\lambda_\nu\})}} \sum_{\bf v} \exp\left[ - \sum_{\nu = 1}^K \lambda_\nu O_\nu ({\bf v})\right] O_\mu ({\bf v})\\
&=& - \sum_{\bf v}{1\over{ {\cal Z}(\{\lambda_\nu\})}}\exp\left[ - \sum_{\nu = 1}^K \lambda_\nu O_\nu ({\bf v})\right] O_\mu ({\bf v})
\\
&=& -\sum_{\bf v}P({\bf v} ) O_\mu ({\bf v})  .
\end{eqnarray}
\end{widetext}
We recognize this as the expectation value of $O_\mu ({\bf v})$ with respect to the probability distribution $P({\bf v})$.  Thus we have
\begin{equation}
{{\partial \ln {\cal Z}(\{\lambda_\nu\})}\over{{\partial \lambda_\mu}} } =  -\langle O_\mu({\bf v})\rangle_P ,\label{thermo_derivs}
\end{equation}
and hence Eq (\ref{maxlike1}) becomes
\begin{equation}
\langle O_\mu({\bf v})\rangle_P  = \langle O_\mu({\bf v})\rangle_{\rm exp}.
\end{equation}
That is, once we have the form of the maximum entropy distribution in Eq (\ref{P_maxent}), maximizing the likelihood of the data with respect to parameters is equivalent to imposing the constraints in Eq (\ref{fixedOs}).

\section{Maximum entropy model for flocks}
\label{maxent-flocks}

Let us now apply the maximum entropy approach to the case of bird flocks. The state  of the system is characterized by the set ${\bf v}\equiv \{ \vec{\mathbf{v}}_1,\vec{\mathbf{v}}_2\cdots, \, \vec{\mathbf{v}}_N\}$  of the individual bird velocities.  As discussed in the main text, we consider observables that measure the local correlations between birds and their neighbors, and the mean and variance of flight speeds.

When we look at a snapshot of the flock, we can identify bird $\rm j$ as being in the neighborhood  of bird $\rm i$ (${\rm j} \in {\cal N}_{\rm i}$) if it is one of the closest $n_c$ neighbors.  Then we measure the mean--square difference in velocity between a bird and those in its neighborhood,
\begin{equation}
Q_{\rm int} = {1\over {2N v_0^2 }}\sum_{{\rm i}=1}^N {1\over {n_c}}\sum_{{\rm j} \in {\cal N}_{\rm i}} |\vec{\mathbf{v}}_{\rm i} - \vec{\mathbf{v}}_{\rm j} |^2 ,
\label{app_Qint}
\end{equation}
where we have normalized by a scale $v_0$ to obtain a dimensionless measure; in solving the model we shall see that it is natural to set this scale equal to the observed mean speed of the birds.  It will be convenient to write this in a slightly different form, so we introduce  matrix $\hat n_{\rm ij} = 1$ if ${\rm j} \in {\cal N}_{\rm i}$ and $\hat n_{\rm ij} =0$ otherwise.  Then we have
\begin{equation}
Q_{\rm int} = {1\over {2N v_0^2 }} {1\over {n_c}} \sum_{{\rm i}=1}^N \sum_{{\rm j} =1}^N\hat n_{\rm ij}  |\vec{\mathbf{v}}_{\rm i} - \vec{\mathbf{v}}_{\rm j} |^2  .
\end{equation}
We notice that the indices $\rm i$ and $\rm j$ appear symmetrically, but the matrix $\hat n_{\rm ij}$ is not symmetric, since ``being in the neighborhood'' is not a symmetrical relationship (if you are my nearest neighbor, I might not be your nearest neighbor).  Only the symmetric part survives the summation, so we can write
\begin{equation}
Q_{\rm int} = {1\over {2N v_0^2 }} {1\over {n_c}} \sum_{{\rm i}=1}^N \sum_{{\rm j} =1}^N n_{\rm ij}  |\vec{\mathbf{v}}_{\rm i} - \vec{\mathbf{v}}_{\rm j} |^2  ,
\end{equation}
where $n_{\rm ij} = (\hat n_{\rm ij} + \hat n_{\rm ji})/2$.

In addition to $Q_{\rm int}$, we chose as observables the mean speed and the mean--square speed across the flock,
\bea
V&=&\frac{1}{N}\sum_{{\rm i}=1}^N v_{\rm i}
\label{v1}\\
V_2&=&\frac{1}{N}\sum_{{\rm i}=1}^N v_{\rm i}^2,
\label{v2-si}
\eea
where $v_{\rm i}=|\vec{{\mathbf v}}_{\rm i}|$ is the speed of bird $\rm i$.

Equation (\ref{app_energy}) tells us that the effective energy function or Hamiltonian for a maximum entropy model is composed of one term for each of the observables whose expectation values we match to the data. Thus we should have
\be
{\cal H}({\bf v}) = \lambda_1 Q_{\rm int} + \lambda_1 V + \lambda_3 V_2 ,
\ee
and the probability distribution
\be
P({\bf v})=\frac{e^{-{\cal H}({\bf v}) }}{\Zt} .
\label{p-maxent}
\ee
It will be useful to absorb factors of $N$ so that the effective energy becomes ``extensive,'' that is proportional (on average) to the number of birds in the flock, while the parameters of the model remain formally independent of $N$.  Similarly, we would like to separate the choice of units for velocity from the dimensionless parameters of our model, so we introduce a scale $v_0$ as in the main text. Thus we write
\be
{\cal H}({\bf v}) = \frac{J}{4v_0^2} \sum_{\rm i,j =1}^N n_{\rm ij} |\vec{\mathbf{v}}_{\rm i} - \vec{\mathbf{v}}_{\rm j} |^2   + \frac{g}{2v_0^2}\sum_{{\rm i}=1}^N v_{\rm i}^2 - {\mu \over {v_0}}\sum_{{\rm i}=1}^N v_{\rm i} .\label{hamilton}\\
\ee
With $P(\mathbf{v}) \propto \exp [ - {\cal H}({\bf v})]$, we obtain Eq (\ref{maxent}) of the main text.

\section{Solving the model}
\label{app:solving}

The first step in using the maximum entropy model is to compute the partition function $\Zt$.  Since the role of $\Zt$ is to enforce normalization, we have
\be
{\cal Z}(J,g,\mu) = \int d{\bf v} \ e^{-{\cal H}(\bf{v})} ,
\label{partfun}
\ee
where $d{\bf v}$ is the volume element in the space of all the (three--dimensional) velocities,
$d{\bf v} = \prod_{\rm i} d^3 \vec{\bf v}_{\rm i}$.

\subsection{Computation with free boundary conditions}
\label{free}

We begin by treating all birds as equivalent, without regard to their location in the interior or on the boundary of the flock, and we return to this below.  It will be useful to think of the velocity as being composed of a speed and a direction, $\vec{{\mathbf v}}_{\rm i} = v_{\rm i} \vec{{\mathbf s}}_{\rm i}$, where $|\vec{{\mathbf s}}_{\rm i}|=1$.  Translating   into these variables, we obtain from  Eq (\ref{hamilton}):
\begin{widetext}
\begin{eqnarray}
{\cal H}({\bf v}) &=& \frac{J}{4v_0^2} \sum_{\rm i,j =1}^N n_{\rm ij} |v_{\rm i} \vec{\mathbf{s}}_{\rm i} - v_{\rm j} \vec{\mathbf{s}}_{\rm j} |^2   
+ \frac{g}{2v_0^2}\sum_{{\rm i}=1}^N v_{\rm i}^2 - {\mu \over {v_0}} \sum_{{\rm i}=1}^N v_{\rm i}\\
&=& \frac{J}{4v_0^2} \sum_{\rm i,j =1}^N n_{\rm ij} \left[ v_{\rm i}^2 - 2 v_{\rm i}v_{\rm j} \vec{\mathbf{s}}_{\rm i} \cdot \vec{\mathbf{s}}_{\rm j} + v_{\rm j}^2\right] 
+ \frac{g}{2v_0^2}\sum_{{\rm i}=1}^N v_{\rm i}^2 -{\mu \over {v_0} }\sum_{{\rm i}=1}^N v_{\rm i}\\
&=& - \frac{J}{2v_0^2}\sum_{\rm i,j =1}^N n_{\rm ij} v_{\rm i}v_{\rm j} \vec{\mathbf{s}}_{\rm i} \cdot \vec{\mathbf{s}}_{\rm j}  
+ {1\over {2v_0^2}} \sum_{{\rm i}=1}^N \left( g + J\sum_{{\rm k}=1}^N n_{\rm ik}\right) v_{\rm i}^2 - {\mu \over {v_0}} \sum_{{\rm i}=1}^N v_{\rm i}  .
\label{H3}
\end{eqnarray}
Notice that the term controlling the mean--square speed now has two contributions, one from the ``direct'' control parameter $g$ and one from the social interactions with neighbors, $\propto J$.

In addition to rewriting the Hamiltonian, we also need to express the volume element $d{\bf v}$ in terms of the new direction and speed variables.  For each bird,
\begin{equation}
d^3 {\bf v}_{\rm i} = v_{\rm i}^2 dv_{\rm i} d^3 \vec{\mathbf{s}}_{\rm i} \delta (|\vec{{\mathbf s}}_{\rm i}| -1) ,
\end{equation}
where the delta function enforces the constraint that $\vec{\mathbf s}_{\rm i}$ is a unit vector, and the factor $v_{\rm i}^2$ is the Jacobian of the transformation.  In the limit that speed fluctuations are small---which they are in the flock---the effect of the Jacobian can always be absorbed into a redefinition of the parameters $\mu$ and $g$, so we drop this term here.  Thus we have
\begin{equation}
\Zt = \int \prod_{{\rm i}=1}^N dv_{\rm i} d^3 \vec{\mathbf{s}}_{\rm i} \delta (|\vec{{\mathbf s}}_{\rm i}| -1) 
\exp\left[  \frac{J}{2v_0^2}\sum_{\rm i,j =1}^N n_{\rm ij} v_{\rm i}v_{\rm j} \vec{\mathbf{s}}_{\rm i} \cdot \vec{\mathbf{s}}_{\rm j}  
- {1\over {2v_0^2}} \sum_{{\rm i}=1}^N \left( g + J \sum_{{\rm k}=1}^N n_{\rm ik}\right) v_{\rm i}^2 + {\mu \over {v_0}} \sum_{{\rm i}=1}^N v_{\rm i}\right]
\end{equation}
\end{widetext}

Now we want to use the fact that fluctuations are small in order to simplify our calculation; we can verify, at the end, that the fluctuations predicted by the model really are small, and hence that our approximations are consistent.  This is a now classical approximation scheme in the theory of magnetism \cite{ref-spinwave}, but we go through the details here in the hopes of making the calculation accessible to a broader audience. 

We can write the speeds as
\begin{equation}
v_{\rm i} = V (1 + \epsilon_{\rm i}) ,
\end{equation}
where $V$ is the mean speed over the flock from Eq (\ref{v1}),
\begin{equation}
V = {1\over N}\sum_{{\rm i}=1}^N v_{\rm i} ,
\end{equation}
and $\epsilon_{\rm i}$ is the fractional fluctuation around this mean; we expect $|\epsilon_{\rm i}| \ll 1$.  Notice that with this definition we have
\begin{equation}
\sum_{{\rm i}=1}^N \epsilon_{\rm i} = 0.
\end{equation}
Transforming from integrating over speeds to integrating over their fluctuations, we have  
\begin{equation}
\prod_{{\rm i}=1}^N dv_{\rm i} = V^N dV \left( \prod_{{\rm i}=1}^N d\epsilon_{\rm i} \right) \delta\left(\sum_{{\rm j}=1}^N \epsilon_{\rm j}\right) .
\label{eps_measure}
\end{equation}

To say that fluctuations in direction are small requires a bit more care.  We can average the unit vectors $\vec{\mathbf s}_{\rm i}$ to obtain the polarization of the flock as in Eq (\ref{Pdef}),
\begin{equation}
\vec{\mathbf{P}} = {1\over N}\sum_{{\rm i}=1}^N \vec{\mathbf{s}}_{\rm i} .
\label{app_P}
\end{equation}
This polarization has a magnitude $P$ and a direction that we will denote by the unit vector $\hat {\mathbf n}$, so that $\vec{\mathbf{P}} = P \hat {\mathbf n}$.  We expect that flight directions of individual birds will be close to $\hat {\mathbf n}$, so we can write
\begin{equation}
\vec{\mathbf{s}}_{\rm i} = s_{\rm i}^L \hat {\mathbf n} + \vec{\ppi}_{\rm i} ,
\label{s_decomp}
\end{equation}
where $\vec{\ppi}_{\rm i}$ is a (small) vector perpendicular to $\hat {\mathbf n}$, and the ``longitudinal'' term $s_{\rm i}^L$ is necessary to be sure that $\vec{\mathbf{s}}_{\rm i}$ remains a unit vector.  As with the $\epsilon_{\rm i}$ above, not all $N$ of these variables are independent, since the definition of the polarization in Eq (\ref{app_P}) requires that
\begin{equation}
P = {1\over N}\sum_{{\rm i}=1}^N s_{\rm i}^L  ,
\ee
and
\be
\sum_{{\rm i}=1}^N \vec{\ppi}_{\rm i}  = 0 .
\ee
\begin{widetext}
Thus we have 
\begin{equation}
\prod_{{\rm i}=1}^N d^3 \vec{\mathbf{s}}_{\rm i} \delta (|\vec{{\mathbf s}}_{\rm i}|-1)  =
\int {{d^2  \vec{\mathbf n}}\over{4\pi}}\int  dP 
\left[\prod_{{\rm i}=1}^N d^2 {\ppi}_{\rm i} ds_{\rm i}^L 
\delta\left( \sqrt{[s_{\rm i}^L ]^2 + |\vec{\ppi}_{\rm i}|^2} - 1\right) \right]
\delta\left( P - {1\over N} \sum_{{\rm i}=1}^N s_{\rm i}^L  \right)
\delta \left(\sum_{{\rm i}=1}^N \vec{\ppi}_{\rm i} \right)
\label{s_measure}
 \end{equation}
 Now, if we substitute into Eq (\ref{H3}), we have
 \begin{equation}
{\cal H}({\bf v}) =
- \frac{JV^2}{2v_0^2}\sum_{\rm i,j =1}^N n_{\rm ij} (1+ \epsilon_{\rm i})(1+\epsilon_{\rm j})\left( s_{\rm i}^L s_{\rm j}^L + \vec{\mathbf {\pi}}_{\rm i} \cdot\vec{\mathbf {\pi}}_{\rm j}  \right)
+ {{V^2}\over {2v_0^2}} \sum_{{\rm i}=1}^N \left( g + {J}\sum_{{\rm k}=1}^N n_{\rm ik}\right) (1+\epsilon_{\rm i})^2  -{{N\mu}\over {v_0}} V
\label{H4}
\end{equation}
Although we have changed variables in a way that makes it easy to make the approximation that fluctuations are small, we haven't actually used this approximation yet in simplifying the Hamiltonian.

We notice that one set of delta functions in Eq (\ref{s_measure}) enforces
\begin{equation}
s_{\rm i}^L = \sqrt{1-|\vec{\mathbf {\pi}}_{\rm i}|^2} \approx 1 - |\vec{\mathbf {\pi}}_{\rm i}|^2 /2 + \cdots ,
\end{equation}
where the approximation is that $ |\vec{\mathbf {\pi}}_{\rm i}|$ is small.  If we substitute this into Eq (\ref{H4}), then to be consistent we should keep only terms up to second order in $ \vec{\mathbf {\pi}}_{\rm i}$ and $\epsilon_{\rm i}$.  The result is
 \begin{eqnarray}
{\cal H}({\bf v}) &=&
- \frac{JV^2}{2v_0^2}\sum_{\rm i,j =1}^N n_{\rm ij} (1+ \epsilon_{\rm i})(1+\epsilon_{\rm j})+{{V^2}\over {2v_0^2}}  \sum_{{\rm i}=1}^N \left( g  + {J}\sum_{{\rm k}=1}^N n_{\rm ik}\right) (1+\epsilon_{\rm i})^2  -{{N\mu}\over {v_0}} V
\nonumber\\
&&\,\,\,\,\,\,\,\,\,\,\,\,\,\,\,\,\,\,\, - \frac{JV^2}{2v_0^2} \sum_{\rm i,j =1}^N n_{\rm ij} \left( - |\vec{\mathbf {\pi}}_{\rm i}|^2 /2 - |\vec{\mathbf {\pi}}_{\rm j}|^2 /2 + \vec{\mathbf {\pi}}_{\rm i} \cdot\vec{\mathbf {\pi}}_{\rm j}  \right) .
\end{eqnarray}
 \end{widetext}
A crucial simplification is that the terms related to speed fluctuations ($\epsilon_{\rm i}$) are decoupled from those related to directional fluctuations ($\vec{\mathbf {\pi}}_{\rm i}$).  Thus we have, as in Eq (\ref{sep1}),
  \begin{equation}
{\cal H}({\bf v}) = {\cal H}_{\rm dir} (\{ \vec{\mathbf {\pi}}_{\rm i} \}) + {\cal H}_{\rm sp} (\{ \epsilon_{\rm i}\}) + E_0(V),
\end{equation}
where $E_0(V)$ is the effective energy when all $\epsilon_{\rm i} = 0$,
\begin{equation}
E_0(V) = N \left( {{gV^2}\over{2 v_0^2}}  - \mu V\right) .
\end{equation}

Collecting terms, and dropping constants independent of $\{ \vec{\ppi}_{\rm i} \}$ and $\{ \epsilon_{\rm i}\}$, we find that
\begin{eqnarray}
{\cal H}_{\rm dir} (\{ \vec{\mathbf {\pi}}_{\rm i} \}) &=&  \frac{JV^2}{2v_0^2}\sum_{\rm i,j =1}^N N_{\rm ij} \vec{\ppi}_{\rm i} \cdot\vec{\ppi}_{\rm j} \label{Hdir_a}\\
{\cal H}_{\rm sp} (\{ \epsilon_{\rm i}\}) &=& \frac{V^2}{2v_0^2}\sum_{\rm i,j =1}^N \left( g \delta_{\rm ij} + J N_{\rm ij}\right) \epsilon_{\rm i}\epsilon_{\rm j}  , \label{Hsp_a}
\end{eqnarray}
where the matrix $N_{\rm ij}$ has the form
\begin{equation}
N_{\rm ij} = - n_{\rm ij} + \delta_{\rm ij} \sum_{{\rm k}=1}^N n_{\rm ik} .
\end{equation}

In trying to compute the partition function, we will need to integrate not just over the ``local'' variables $\{\epsilon_{\rm i} , \vec{\ppi}_{\rm i}\}$, but also---as can be seen from the volume elements in Eqs (\ref{eps_measure}) and (\ref{s_measure})---over the global variables $V$, $P$, and $\hat{\mathbf n}$.  The integral over the direction of polarization is simple because there is no dependence of the integrand on $\hat{\mathbf n}$; this is a consequence of the overall rotational invariance in our formulation of the problem.  The integral over the magnitude of the polarization is also simple, since the delta function just gives us 
\begin{equation}
P = {1\over N}\sum_{{\rm i}=1}^N s_{\rm i}^L \approx  1 - {1\over {2N}} \sum_{{\rm i}=1}^N |\vec{\mathbf {\pi}}_{\rm i}|^2 .
\end{equation}

The integral over $V$ is more interesting, since the $V$ dependence of the integrand is dominated by $E_0(V)$.  Thus we need to do an integral of the form
\begin{equation}
{\cal Z}_V \approx \int  dV\, e^{-E_0(V)} .
\end{equation}
The key point is that $E_0 \propto N$, and so the integrand is very sharply peaked around some $V_*$.  But the average of $V$ is one of the quantities that we are fixing from the data, so we must have $V_* = \langle V\rangle _{\rm exp}$, and this serves to set the parameter $\mu$, as explained in the main text.  Importantly, the factor of $N$ insures that the variations in $V$ around $V_*$ will be very small in large flocks, and hence we can replace $V \rightarrow V_* = \langle V\rangle_{\rm exp}$ everywhere else in our calculations.  We are also free to choose the scale $v_0 = \langle V\rangle_{\rm exp}$, and then we can simplify
\begin{eqnarray}
{\cal H}_{\rm dir} (\{ \vec{\mathbf {\pi}}_{\rm i} \}) &=&  \frac{J}{2}\sum_{\rm i,j =1}^N N_{\rm ij} \vec{\ppi}_{\rm i} \cdot\vec{\ppi}_{\rm j} ,
\label{Hdir}\\
{\cal H}_{\rm sp} (\{ \epsilon_{\rm i}\}) &=& \frac{1}{2}\sum_{\rm i,j =1}^N \left( g  \delta_{\rm ij} + J N_{\rm ij}\right) \epsilon_{\rm i}\epsilon_{\rm j}  .
\label{Hsp}
\end{eqnarray}

This separation of direction and speed variables in the Hamiltonian means that the partition function can be factorized,
\begin{equation}
\Zt \propto {\cal Z}_{\rm dir} (J) {\cal Z}_{\rm sp}(J , g) e^{Ng/2},
\label{Zfac}
\end{equation}
where
\begin{eqnarray}
{\cal Z}_{\rm dir} (J) &=& \int \left[\prod_{{\rm i}=1}^N d^2 {\ppi}_{\rm i}\right] \delta \left(\sum_{{\rm i}=1}^N \vec{\ppi}_{\rm i} \right) e^{- {\cal H}_{\rm dir} (\{ \vec{\mathbf {\pi}}_{\rm i} \})}
\nonumber\\
&&\label{Zdir}
\\
{\cal Z}_{\rm sp}(J , g) &=&\int \left[\prod_{{\rm i}=1}^N d\epsilon_{\rm i}\right]\delta\left(\sum_{{\rm j}=1}^N \epsilon_{\rm j}\right) e^{- {\cal H}_{\rm sp} (\{ \epsilon_{\rm i} \})} .
\nonumber\\
&&\label{Zsp}
\end{eqnarray}

Now we have to do the integrals in Eqs (\ref{Zdir}) and (\ref{Zsp}), but these are not so difficult because they are Gaussians.  The behavior of these integrals is determined the structure of the matrix $N_{\rm ij}$.  To understand this structure, imagine that the birds are in a line, and the relevant neighborhood is just the two nearest neighbors along the line.  Then we can see that $N_{\rm ij}$ is the discrete approximation to the (negative) second derivative along the line.  In higher dimensions this becomes the Laplacian operator, and so $N_{\rm ij}$ is called a Laplacian matrix.  As with the negative Laplacian, the eigenvalues $\{\Lambda_{\rm a}\}$ of $N_{\rm ij}$ are positive, except for the smallest one, which exactly zero ($\Lambda_1 = 0$).  If we define the eigenvectors of $N_{\rm ij}$ by $w_{\rm i}^{\rm a}$ such that
\begin{equation}
\sum_{{\rm j}=1}^N N_{\rm ij} w_{\rm j}^{\rm a} = \Lambda_{\rm a} w_{\rm i}^{\rm a},
\label{eigdef}
\end{equation}
then the  eigenvector associated with the zero eigenvalue is the ``uniform'' mode, $w_{\rm i}^{\rm 1} = {\rm constant}$.  But displacements along this direction are fixed to zero by the delta functions that appear in the integrals of Eqs (\ref{Zdir}) and (\ref{Zsp}), and this is crucial for doing the integrals.

We recall that, for a general $N\times N$ matrix $M_{\rm ij}$,
\begin{widetext}
\begin{equation}
\int d^N x \exp\left( - {1\over 2} \sum_{{\rm i,j}=1}^N x_{\rm i} M_{\rm ij} x_{\rm j}\right) = \left[ {{(2\pi)^N}\over{\det M}}\right]^{1/2} \propto \exp\left( - {1\over 2} \sum_{{\rm a} =1}^N \ln [\lambda_{\rm a}(M)] \right),
\label{gaussint}
\end{equation}
where $\lambda_{\rm n}(M)$ are the eigenvalues of $M$.  In the case of ${\cal Z}_{\rm sp}$, we have
\begin{equation}
{\cal Z}_{\rm sp}(J , g) = \int \left[\prod_{{\rm i}=1}^N d\epsilon_{\rm i}\right]\delta\left(\sum_{{\rm j}=1}^N \epsilon_{\rm j}\right)\exp\left[ -{1\over 2} \sum_{{\rm i,j}=1}^N \epsilon_{\rm i} (g\delta_{\rm ij} + J N_{\rm ij}) \epsilon_{\rm j}\right] .
\end{equation}
\end{widetext}
 The relevant matrix is now $M_{\rm ij} = g\delta_{\rm ij} + J N_{\rm ij}$, and the eigenvalues are $\lambda_{\rm a}(M) = g + J \Lambda_{\rm a}$, where again $\Lambda_{\rm a}$ are the eigenvalues of the Laplacian matrix $N_{\rm ij}$.  We note that the integral runs over $N$ dimensions, but the delta function fixes one combination of the $\{\epsilon_{\rm i}\}$ to be zero, and as noted above this combination is parallel to the first eigenvector.  So, up to constant factors, the effect of the delta function is to exclude the first (zero) eigenvalue from the sum in Eq (\ref{gaussint}), so that
 \begin{equation}
{\cal Z}_{\rm sp}(J , g) \propto \exp\left( - {1\over 2} \sum_{{\rm a} =2}^N \ln [g + J \Lambda_{\rm a}] \right) .
\label{Zsp_free}
\end{equation}

Since the effective Hamiltonian for speed fluctuations in Eq (\ref{Hsp}) is a quadratic function of the $\{\epsilon_{\rm i}\}$,  the probability distribution of the speed fluctuations is  Gaussian,
\begin{equation}
P(\{ \epsilon_{\rm i}\}) = {1\over{{\cal Z}_{\rm sp}(J,g)}} \delta\left(\sum_{{\rm j}=1}^N \epsilon_{\rm j}\right)\exp\left[ -{1\over 2} \sum_{{\rm i,j}=1}^N \epsilon_{\rm i} M_{\rm ij}  \epsilon_{\rm j}\right] .
\end{equation}
Thus we can calculate the correlations between the values of $\epsilon$ for different birds $\rm i$ and $\rm j$ in a standard way:  we rotate our coordinates into the eigenvectors of the matrix $M_{\rm ij}$, we note that in this basis fluctuations along each coordinate are independent with variance $1/\Lambda_{\rm n}(M)$, and then to recover the correlations in the original basis we rotate back.  Again we have to be careful to respect the delta function, which serves to eliminate the fluctuations along $w_{\rm i}^1$.  The end result is that
\begin{equation}
\langle \epsilon_{\rm i}\epsilon_{\rm j}\rangle = \sum_{{\rm a}=2}^N {{w_{\rm i}^{\rm a} w_{\rm j}^{\rm a}}\over{g +J \Lambda_{\rm a}}} .
\label{eps_corr1}
\end{equation}
This result, or more precisely its generalization to the case where we treat the birds on the boundary of the flock separately, Eq (\ref{eps_corr2}),  is the basis for our prediction of the speed correlations as a function of the distance between birds, in Fig \ref{fig:correlations}c.

We can carry through the same calculation for the direction fluctuations.  The only differences are  that the vector $\vec{\ppi}_{\rm i}$ has two components, so there are twice as many variables, and that the matrix which controls the fluctuations is now simple $M_{\rm ij} = J N_{\rm ij}$.  The results are 
\begin{equation}
{\cal Z}_{\rm dir}(J ) \propto \exp\left( -  {{d-1}\over{2}}\sum_{{\rm a} =2}^N \ln [J \Lambda_{\rm a}] \right) ,
\label{Zdir_free}
\end{equation}
and
\begin{equation}
\langle \vec{\ppi}_{\rm i}\cdot \vec{\ppi}_{\rm j}\rangle = (d-1) \sum_{{\rm a}=2}^N {{w_{\rm i}^{\rm a} w_{\rm j}^{\rm a}}\over{J \Lambda_{\rm a}}} ,
\label{pi_corr1}
\end{equation}
where we give the result for motion in $d$ dimensions; here $d=3$.

As noted at the end of Appendix \ref{maxentapp}, imposing the constraint that expectation values of observables in our model be equal to those found in the data is equivalent to maximum likelihood inference.  Thus, to complete our calculation and find the parameters of our model, we should compute the probability of the data in the model, as function of the parameters $J$, $g$, and $n_c$.  Putting together the results in this section, we can write the log of the full probability distribution as
\begin{widetext}
\begin{eqnarray}
\Phi \equiv \ln P({\rm data}|{\rm model}) &=& -\ln {\cal Z} - \langle {\cal H}({\bf v})\rangle_{\rm exp}\\
&=& -\ln {\cal Z}_{\rm dir}(J) - \ln {\cal Z}_{\rm sp}(J,g) -  {\bigg\langle}\frac{J}{4V^2}  \sum_{\rm i,j =1}^N n_{\rm ij} |\vec{\mathbf{v}}_{\rm i} - \vec{\mathbf{v}}_{\rm j} |^2 {\bigg\rangle}_{\rm exp}  - {\bigg\langle}\frac{g}{2V^2}\sum_{{\rm i}=1}^N\left( v_{\rm i}- V\right) ^2 {\bigg\rangle}_{\rm exp} \\
&=&  \sum_{{\rm a} =2}^N \ln [J \Lambda_{\rm a}]  + {1\over 2} \sum_{{\rm a} =2}^N \ln [g + J \Lambda_{\rm a}] - N {{J n_c}\over 2} \langle Q_{\rm int}\rangle_{\rm exp} -N \frac{g}{2} \langle \sigma^2\rangle_{\rm exp} ,
\label{Phi_free}
\end{eqnarray}
\end{widetext}
where $\langle \cdots \rangle$ denotes an average over the data,  we identify $Q_{\rm int}$ from Eq (\ref{Qint}) of the main text, and $\sigma^2$ is the fractional variance of individual birds' speeds around the flock mean.  

The result for $\Phi$ in Eq (\ref{Phi_free}) is simple enough that we can maximize to give explicit equations that determine the parameters.  Thus
\begin{eqnarray}
{{\partial \Phi}\over{\partial g}}&=& 0\\
\Rightarrow {1\over N} \sum_{{\rm a} =2}^N {1\over {g + J \Lambda_{\rm a}}} &=& \langle \sigma^2\rangle_{\rm exp} ,
\label{gfreecond}
\end{eqnarray}
and similarly
\begin{eqnarray}
{{\partial \Phi}\over{\partial J}}&=& 0\\
\Rightarrow {{(N-1)}\over J} + {1\over 2} \sum_{{\rm a} =2}^N {{\Lambda_{\rm a}} \over {g + J \Lambda_{\rm a}}} &=& N {{n_c}\over 2} \langle Q_{\rm int}\rangle_{\rm exp} 
\nonumber\\
&&\\
d \left( 1 - {1\over N}\right) - g \langle \sigma^2\rangle_{\rm exp} &=& J n_c  \langle Q_{\rm int}\rangle_{\rm exp} .
\nonumber\\
&&\label{Jfreecond}
\end{eqnarray}
Finally, we can substitute the solutions to these equations, $J^*$ and $g^*$, back into Eq (\ref{Phi_free}) and maximize with respect to $n_c$, as in Fig \ref{fig:parameters}c.

\subsection{Computation with fixed boundary conditions}
\label{fixed}

 So far, we have assumed free boundary conditions, corresponding to the ideal situation where speed and orientations of all individuals in a  flock can fluctuate in the same manner, exploring the whole accessible space of possible fluctuations, given the interaction between birds. In natural flocks this is not very realistic: individuals on the boundary are constantly subject to environmental stimuli, so that they will adjust their direction and speed not only in response to neighboring birds, but also in response to external cues. To cope with this fact, we now perform the computation of the partition functions and of the likelihood using ``fixed boundary conditions,'' where the velocities of the birds on the boundary of the flock are held fixed at their observed values. We note that for large systems, such as the flocks we are considering, boundary individuals are a negligible fraction of all individuals. As  discussed more fully  in Ref \cite{bialek+al_12}, the values of the inferred parameters do not change much with changing the boundary conditions. Fixed boundary conditions are however necessary to adequately take into account the effects of boundary on the correlations.

To perform the computations with fixed conditions on the border, it is convenient to divide the birds in two groups: internal birds ${\rm i,j} \in \cI$ and birds belonging to the border ${\rm a,b} \in \cB$. Then, Eq  (\ref{hamilton}) becomes
\begin{widetext}
\be
{\cal H}({\bf v}) = \frac{J}{2v_0^2} \sum_{{\rm i,j}\in \cI} \left( N_{\rm ij} + \frac{g}{J} \delta_{\rm ij}\right) \Vi \cdot \Vj      - {J \over {v_0}} \sumI \Hi \cdot \Vi + \cH_{\cB}(J,g) - {\mu\over{v_0}} \sum_{{\rm i}=1}^N v_{\rm i}  ,
\label{hamiltonBis}
\ee
\end{widetext}
where
\bea
\Hi &=& {1\over {v_0}} \sumA n_{\rm ia} \Va  \\
\cH_{\cB}(J,g) &=& \frac{J}{2v_0^2} \sumAB \lx(N_{\rm ab} + \frac{g}{J} \delta_{\rm ab}\rx) \Va \cdot \Vb  .
\eea
We can see from these expressions that holding velocities  ${\vec {\mathbf v}}_{\rm a}$ fixed on the border of the flock is equivalent to considering a flock in presence of a field  $\Hi$ acting on those birds who see the border birds as their neighbors.  Note that birds deep in the interior do not couple directly to the field, but may feel its influence if it propagates through the flock.  It will be useful to decompose these fields in relation to the mean flight direction $\hat{\mathbf n}$, as in Eq (\ref{s_decomp}),
\begin{equation}
\Hi = h_{\rm i}^L \hat{\mathbf n} + \Hi^\perp .
\end{equation}

The computation of the partition function now proceeds exactly as in the previous subsection. The only difference is that integrations must now be performed on internal variables only; the algebra is slightly more complicated, but the conceptual are the same. Corresponding to Eq (\ref{Zfac}) we have  
\be
{\cal Z}(J,g;n_c) = e^{ -\cH_{\cB}(J,g)} {\cal Z}_{\rm dir}(J) {\cal Z}_{\rm sp} (J, g) e^{Ng/2},
\ee
and in place of Eqs (\ref{Zdir}) and (\ref{Zsp}) we have
\bea
{\cal Z}_{\rm dir}(J) &=& 
\int \left[\prod_{{\rm i}\in {\cal I}} d^2 {\ppi}_{\rm i}\right] \delta \left(\sum_{{\rm i}=1}^N \vec{\ppi}_{\rm i} \right)
e^{-{\cal H}_{\rm dir}(\{\vec{\ppi}_{{\rm i}\in \cI}\})}
\nonumber\\
&&  
\label{zor-border}  \\
{\cal Z}_{\rm sp}(J , g) &=&\int \left[\prod_{{\rm i}\in {\cal I}} d\epsilon_{\rm i}\right]\delta\left(\sum_{{\rm j}=1}^N \epsilon_{\rm j}\right) e^{- {\cal H}_{\rm sp} (\{ \epsilon_{{\rm i}\in {\cal I}} \})} ,
\nonumber\\
&&
\label{zsp-border}
\eea
where we note that the integration is only over  internal variables, but the delta function constraints involve all the variables.  As in the case of free boundaries, we first integrate over global variables, which has the effect of pinning the mean velocity to its observed value, and then we can choose the scale $v_0 = \langle V \rangle_{\rm exp}$, simplifying all the expressions. The reduced Hamiltonians for the internal variables, analogs of Eqs (\ref{Hdir}) and (\ref{Hsp}), then become
\bea
{\cal H}_{\rm dir}(\{\vec{\ppi}_{{\rm i}\in \cI}\}) &=& 
\frac{J}{2} \sumIJ N_{\rm ij}   \vec{\ppi}_{\rm i} \cdot \vec{\ppi}_{\rm j}   - J  \sumI \Hi^\perp \cdot \vec{\ppi}_{\rm i}
\nonumber\\
&&\\
{\cal H}_{\rm sp} (\{ \epsilon_{{\rm i}\in {\cal I}} \}) &=& \frac{J}{2} \sumIJ \lx( N_{\rm ij} + \frac{g}{J}   \delta_{\rm ij}\rx) \Ei \Ej   - J  \sumI b_{\rm i} \Ei  ,
\nonumber\\
&&
\eea
where 
\begin{equation}
b_{\rm i}=h_{\rm i}^L-\sum_{{\rm a}\in{\cal B}} n_{\rm ia}=\sum_{{\rm a}\in{\cal B}} n_{\rm ia}\eps_{\rm a}\end{equation}
 is the fluctuating part of the longitudinal component of border field. 
 
 Although we have same matrix $N_{\rm ij}$ in these equations as in the previous section, the indices $\rm ij$ are restricted to the interior of the flock, and on this restricted space the matrix has different properties.  To remind us of this fact, it  is convenient to introduce the two matrices $A_{\rm ij} = N_{\rm ij} $ and $B_{\rm ij} = N_{\rm ij} + (g/J) \delta_{\rm ij}$, with indices that refer only to birds internal to the flock, ${\rm i} \in {\cal I}$.  Then the partition functions that we need to evaluate are again Gaussian integrals, controlled by the properties of these matrices.  We find, corresponding to Eqs (\ref{Zsp_free}) and (\ref{Zdir_free}), 
\begin{widetext}
\bea
\ln {\cal Z}_{\rm dir}(J) &=& \frac{J}{2} \sumIJ (A^{-1})_{\rm ij} \Hi^\perp \cdot \Hj^\perp -{{d-1}\over2}\lx(N_\cI - 1\rx)\ln{\lx(J\rx)} 
- {{d-1}\over2}\ln\left[\sumIJ (A^{-1})_{\rm ij}\right]
\nonumber\\
&&\,\,\,\,\,\,\,\,\,\,\,\,\,\,\,\,\,\,\,\,\,\,\,\,\,\,\,\,\,\,\,\,\,\,\,\,\,\,
- {{d-1}\over2}\ln\det A   - \frac{J}{2} \lx|  \sumA \Paia + \sumIJ (A^{-1})_{\rm ij} \Hi^\perp \rx|^2 \frac{1}{\sumIJ (A^{-1})_{\rm ij}} ,
\eea
and
\bea
\ln {\cal Z}_{\rm sp}(J , g) &=&  \frac{J}{2} \sumIJ (B^{-1})_{\rm ij} b_{\rm i} b_{\rm j} -\frac{1}{2}\lx(N_\cali - 1\rx)\log{\lx(J\rx)}  - \frac{1}{2}\log\left[\sumIJ (B^{-1})_{\rm ij}\right] 
\nonumber\\
&&\,\,\,\,\,\,\,\,\,\,\,\,\,\,\,\,\,\,\,\,\,\,\,\,\,\,\,\,\,\,\,\,\,\,\,\,\,\,
- \frac{1}{2} \ln\det B  - \frac{J}{2} \lx|  \sumA \Ea + \sumIJ (B^{-1})_{\rm ij} b_{\rm i} \rx|^2 \frac{1}{\sumIJ (B^{-1})_{\rm ij}} .
\eea
\end{widetext}
Similarly, the probability distributions of the variables $\{\epsilon_{\rm i} , \vec{\ppi}_{\rm i}\}$ again are Gaussian, and we can find, by analogy with Eqs (\ref{eps_corr1}) and (\ref{pi_corr1}), the correlation functions.  One new feature is that birds in the interior can have nonzero averages of these fluctuations, since they are responding to the birds on the boundary.  Instead of rotating to the basis of eigenvectors, it is useful to define the matrices
\bea
\widetilde{A}_{\rm ij} &=& (A^{-1})_{\rm ij} - \frac{\sum_{{\rm l}\in\cI}(A^{-1})_{\rm il} \sum_{{\rm m}\in\cI}(A^{-1})_{\rm jm}}{\sum_{{\rm l, m}\in\cI}(A^{-1})_{\rm lm}}  ,\\
\widetilde{B}_{\rm ij} &=& (B^{-1})_{\rm ij} - \frac{\sum_{{\rm l}\in\cI}(B^{-1})_{\rm il} \sum_{{\rm m}\in\cI}(B^{-1})_{\rm jm}}{\sum_{{\rm l,m}\in\cI}(B^{-1})_{\rm lm}} .
\eea
Then we find the mean directional fluctuation and the correlations in these fluctuations to be
\bea
\langle \vec{\ppi}_{\rm i} \cdot \vec{\ppi}_{\rm j} \rangle & = &\frac{d-1}{J} \widetilde{A}_{\rm ij} +\langle \vec{\ppi}_{\rm i} \rangle \cdot \langle \vec{\ppi}_{\rm j}\rangle , \label{pi_corr2}\\
 \langle \vec{\ppi}_{\rm i} \rangle& = &\sumJ \widetilde{A}_{\rm ij}  {\Hj^\perp}  - \frac{\sumJ (A^{-1})_{\rm ij}}{\sum_{{\rm l , m} \in \cI} (A^{-1})_{\rm lm}}\sumA \vec{\ppi}_{\rm a} .
\label{aveori}\nonumber\\
&&
\eea
Similarly, we find the mean speed fluctuation and correlations to be
\bea
 \langle \Ei \cdot \Ej \rangle & = &\frac{1}{J} \widetilde{B}_{\rm ij} +\langle \Ei \rangle \cdot \langle \Ej \rangle , \label{eps_corr2}\\
 \langle \Ei \rangle &=& 
\sumJ \widetilde{B}_{\rm ij} b_{\rm j} - \frac{\sumJ (B^{-1})_{\rm ij}}{\sum_{{\rm l , m}\in \cI} (B^{-1})_{\rm lm}}\sumA \Ea .
\label{avespe}
\eea
The correlation functions that we present in Figs \ref{fig:correlations} and \ref{fig:multi-g} are based on these expressions.

Finally, we need to find the conditions that set the values of the parameters. By analogy with Eqs (\ref{gfreecond}) and (\ref{Jfreecond}), we find  
\begin{widetext}
\bea
\frac{1}{J} \lx(d \frac{N^\cI -1}{N} - g \langle \sigma^2\rangle_{\rm exp}\rx) &=&
 n_c  \langle Q_{\rm int}\rangle_{\rm exp}   
+\frac{1}{N }\sumIJ (A^{-1})_{\rm ij} \Hi^\perp \cdot \Hj^\perp +\frac{1}{N}\sumIJ (B^{-1})_{\rm ij} b_{\rm i} b_{\rm j}  
\nonumber\\
&&\,\,\,\,\,\,\,\,\,\,\, 
- \frac{1}{N}\lx|\sumA \vec{\ppi}_{\rm a} + \sumIJ (A^{-1})_{\rm ij} \Hj^\perp \rx|^2 \frac{1}{\sumIJ (A^{-1})_{\rm ij}} \nonumber\\
&&\,\,\,\,\,\,\,\,\,\,\,\,\,\,\,\,\,\,\,\,\,\,   
- \frac{1}{N}\lx(\sumA \Ea + \sumIJ (B^{-1})_{\rm ij} b_{\rm j}\rx)^2 \frac{1}{\sumIJ (B^{-1})_{\rm ij}}
\nonumber\\
&&\,\,\,\,\,\,\,\,\,\,\,\,\,\,\,\,\,\,\,\,\,\,\,\,\,\,\,\,\,\,\,\,\,    
 -\frac{1}{2N\langle V\rangle_{\rm exp}^2}\sum_{{\rm a,b} \in {\cal B} } n_{\rm ab} | \vec{\bf v}_{\rm a} - \vec{\bf v}_{\rm b}|^2 + \frac{g}{NJ\langle V\rangle_{\rm exp}^2} \sum_{{\rm a} \in {\cal B}} (v_{\rm a} - \langle V\rangle_{\rm exp})^2  ,\label{eqJb}
\eea
and 
\be
\langle \sigma^2\rangle_{\rm exp} = \frac{1}{NJ} \sumI \widetilde{B}_{\rm ii} + \frac{1}{N} \sumA \Ea^2 + 
 \frac{1}{N} \frac{ \sumIJ (B^{-2})_{\rm ij} }{\sumIJ (B^{-1})_{\rm ij}}  \lx( \sumA \Ea \rx)^2 \label{eqgb}
\ee
Finally, the optimal value of $n_c$ can be found by maximizing the log-likelihood
\be
\Phi(J,g;n_c)=-\ln{\cal Z}_{\rm dir} (J)-\ln {\cal Z}_{\rm sp} (J,g) +{\cal H}_{\cal B}(J,g) - \frac{J n_c N}{2} \langle Q_{\rm int}\rangle_{\rm exp} - N\frac{g}{2}\langle \sigma^2\rangle_{\rm exp} ,
\label{likelihood_border}
\ee
where we substitute for  $J$ and $g$  the ($n_c$ dependent)  solutions of Eqs (\ref{eqJb}) and(\ref{eqgb}). 
An example of the likelihood as a function of $n_c$ is given in the main text.
\end{widetext}

\section{Goldstone modes and the continuum limit}
\label{app:goldstone}

In this Appendix we would like to make more explicit some of the mathematics behind the intuitions described in Section \ref{sec:intuition} of the main text.    Our discussion is for the case (Appendix \ref{free}) with free boundary conditions.

We start by looking at the effective Hamiltonian for the directional variables $\{ \vec{\mathbf {\pi}}_{\rm i} \}$, in Eq (\ref{Hdir}),
\begin{equation}
{\cal H}_{\rm dir} (\{ \vec{\mathbf {\pi}}_{\rm i} \}) =  \frac{J}{2}\sum_{\rm i,j =1}^N N_{\rm ij} \vec{\ppi}_{\rm i} \cdot\vec{\ppi}_{\rm j}  .
\nonumber
\end{equation}
As explained in the discussion leading up Eq (\ref{eigdef}), the matrix $N_{\rm ij}$ has a zero eigenvalue, but in fact the whole eigenvalue spectrum has a special structure.  To see this, it is useful to imagine that the birds are arranged along a line, and that the neighborhood is only the very nearest neighbor.  Then we can label the birds by $\rm n$, and the bird $n + 1$ is the neighbor of bird $\rm n$; we can rearrange the terms in the sum to give
\begin{equation}
{\cal H}_{\rm dir} (\{ \vec{\mathbf {\pi}}_{\rm i} \}) =  \frac{J}{2}\sum_{\rm n =1}^N |  \vec{\ppi}_{\rm n} - \vec{\ppi}_{{\rm} n +1}  |^2 .
\end{equation}
Now suppose that the direction of flight varies only very slowly, so that we can picture a continuous function of position $x$ in the flock, despite the fact that the birds are located at discrete positions $x_{\rm n} = {\rm n} r_c$, where $r_c$ is the typical distance between the nearest birds.  Then we have $\vec{\mathbf {\pi}}(x)$, and
\begin{equation}
{\cal H}_{\rm dir} (\{ \vec{\ppi}_{\rm i} \}) \approx  \frac{Jr_c^2}{2}\sum_{\rm n =1}^N  {\bigg |}  {{\partial \vec{\ppi} (x)}\over{\p x}}  {\bigg |}^2 .
\end{equation}
Since we are assuming that variations are smooth, we can turn the sum into an integral,
\begin{equation}
{\cal H}_{\rm dir} (\{ \vec{\ppi}_{\rm i} \}) =  \frac{Jr_c^2}{2}\rho \int dx\,  {\bigg |}  {{\partial \vec{\ppi}(x)}\over{\p x}}  {\bigg |}^2 ,
\end{equation}
where $\rho$ is the density of birds along the line.
If we do the same calculation not with birds along a line, but on a regular lattice in three--dimensional space, we find
\begin{equation}
{\cal H}_{\rm dir} (\{ \vec{\mathbf {\pi}}_{\rm i} \}) =  \frac{Jn_c r_c^2}{2}\rho \int d^3x\,  { |} \nabla \vec{\ppi}(x)  { |}^2 ,
\label{pi_cont}
\end{equation}
where we also include the more realistic possibility that the ``neighborhood'' is not just one neighbor but a group of $n_c$ neighbors.

The crucial point about Eq (\ref{pi_cont}) is that if we consider variations in flight direction on a scale $\ell$, such as $\vec{\ppi}(x) \sim A \sin(2\pi x/\ell)$, then we have ${\cal H}_{\rm dir} \propto A^2 /\ell^2$.  Thus, as the length scale of variations becomes large ($\ell \rightarrow\infty$), the ``stiffness'' which resists the variations goes to zero.  This vanishing stiffness at long wavelengths is the signature of a ``Goldstone mode,'' which arises because the original model allowed flight in any direction, but the actual state of the flock breaks this symmetry by selecting a particular direction \cite{stat-mech}.

If the stiffness that opposes variations (in the Hamiltonian) goes down, then the variance of these fluctuations (in the probability distribution) goes up.  Thus in the presence of Goldstone modes we will see a large variance of fluctuations corresponding to variations over long length scales.  In other words, we will see long--ranged correlations.  It is important the these are not just ``long ranged,'' but they are genuinely scale--free. To see this it is useful to remember some mathematical facts about Gaussian random functions (see, for example, Appendix A.2 of Ref \cite{bialek_12}).  

Suppose that we have a function $\phi(x)$, with zero mean.  If all points $x$ are equivalent, we can characterize the statistics of fluctuations in $\phi(x)$ using the correlation function,
\begin{equation}
C_\phi (x - x') = \langle \phi(x) \phi(x')\rangle.
\end{equation}
It is also useful to consider the Fourier transform of the correlation function, the power spectrum,
\begin{equation}
S_\phi(k) = \int dx \, e^{+ikx} C_\phi (x) .
\end{equation}
Importantly, we can write the entire probability distribution for the functions $\phi (x)$ using the power spectrum,
\begin{equation}
P[\phi(x)] = {1\over Z} \exp\left[ - {1\over 2}\int {{dk}\over{2\pi}} {{|\tilde\phi (k)|^2}\over{S_\phi(k)}} \right] ,
\label{gauss_spec}
\end{equation}
where
\begin{equation}
\tilde\phi (k) = \int dx \,e^{+ikx} \phi (x)
\end{equation}
is the Fourier transform of the function $\phi(x)$.

Since we have $P\propto \exp[-{\cal H}]$,  Eq (\ref{pi_cont}) tells us that
\begin{equation}
P[\vec{\ppi} (x)] = {1\over Z} \exp\left[ - \frac{Jn_c r_c^2}{2}\rho \int d^3x\,  {
 |} \nabla \vec{\ppi}(\vec{\mathbf {x}})  { |}^2 \right] .
 \label{Ppi_cont1}
\end{equation}
We can also write this in terms of the Fourier transforms,
\begin{equation}
\tilde{\ppi} (\vec{\mathbf {k}}) = \int dx \, e^{+i\vec{\mathbf {k}}\cdot \vec{\mathbf {x}}} \vec{\ppi}(\vec{\mathbf {x}}) , 
\end{equation}
and then Eq (\ref{Ppi_cont1}) becomes
\begin{equation}
P[\vec{\ppi} (x)] = {1\over Z} \exp\left[ - \frac{Jn_c r_c^2}{2}\rho \int {{d^3 k}\over{(2\pi)^3}} \,  
{|} \vec{\mathbf {k}}{ |}^2 |\tilde{\ppi} (\vec{\mathbf {k}})|^2 \right] .
 \label{Ppi_cont2}
\end{equation}
But now we can read off the power spectrum, by comparing Eqs (\ref{Ppi_cont2}) and (\ref{gauss_spec}); we see that
\begin{equation}
S_{\pi} (\vec{\mathbf {k}} ) = {1\over {Jn_c r_c^2\rho}} \cdot {1\over {{|} \vec{\mathbf {k}}{ |}^2}} .
\label{Spi}
\end{equation}
If we transform back to give the correlation function, we have
\begin{eqnarray}
C_{\pi} (\vec{\mathbf {x}})  &=& \int  {{d^3 k}\over{(2\pi)^3}} \, e^{-i\vec{\mathbf {k}}\cdot \vec{\mathbf {x}}} S_{\pi} (\vec{\mathbf {k}} ) \\
&=& {1\over {Jn_c r_c^2\rho}} \int  {{d^3 k}\over{(2\pi)^3}} \, e^{-i\vec{\mathbf {k}}\cdot \vec{\mathbf {x}}} 
 {1\over {{|} \vec{\mathbf {k}}{ |}^2}} .
 \label{pi_corr1}
 \end{eqnarray}
The key point about this result is that there is nothing in the integral to set a characteristic scale for $\vec{\mathbf {x}}$.  In fact, if we double the value of $|\vec{\mathbf {x}}|$ we make up for this by cutting the value of $|\vec{\mathbf {k}}|$ in half so that $\vec{\mathbf {k}}\cdot \vec{\mathbf {x}}$ stays fixed, but since we are integrating over all possible values of $\vec{\mathbf {k}}$, all that happens is that the whole integral is reduced by a factor of two.  This dimensional analysis argument tells us that
\begin{equation}
C_{\pi} (\vec{\mathbf {x}})  \propto {1\over {|\vec{\mathbf {x}}|}} .
\end{equation}
This is a ``power--law'' decay of correlations with distance (here the power is 1), and {\em it has no characteristic scale.}  Thus, scale--free correlations in directional fluctuations are a consequence of the Goldstone modes.

The predictions for speed fluctuations are very different than for directional fluctuations.  In taking the limit of smooth, continuous variations for directional variations, we found
\begin{eqnarray}
{\cal H}_{\rm dir} (\{ \vec{\ppi}_{\rm i} \}) &=&  \frac{J}{2}\sum_{\rm i,j =1}^N N_{\rm ij} \vec{\ppi}_{\rm i} \cdot\vec{\ppi}_{\rm j}\nonumber\\
&\rightarrow&  \frac{Jn_c r_c^2}{2}\rho \int d^3x\,  { |} \nabla \vec{\ppi}(x)  { |}^2.
\end{eqnarray}
The same argument for speed fluctuations starts with Eq (\ref{Hsp_a}), and gives
\begin{eqnarray}
{\cal H}_{\rm sp} (\{ \epsilon_{\rm i}\}) &=& \frac{1}{2}\sum_{\rm i,j =1}^N \left( {{gV^2}\over{v_0^2}} \delta_{\rm ij} + J N_{\rm ij}\right) \epsilon_{\rm i}\epsilon_{\rm j}\nonumber\\
&\rightarrow& \frac{1}{2}\rho \int d^3x\,  \left[ Jn_c r_c^2{ |} \nabla \epsilon (\vec{\mathbf {x}})  { |}^2
+ g \epsilon^2 (\vec{\mathbf {x}})\right]  \nonumber\\
&&\\
&=&
\frac{1}{2}\rho \int {{d^3 k}\over{(2\pi)^3}}  \left[ Jn_c r_c^2 {|} \vec{\mathbf {k}}{ |}^2 + g \right] 
|\tilde \epsilon (\vec{\mathbf {k}})  { |}^2 ,
\nonumber\\
&&
\end{eqnarray}
where in the last step we transform to the Fourier representation.    By the same argument that leads to Eq (\ref{Spi}), we recognize the predicted power spectrum for fluctuations in the speed,
\begin{equation}
S_\epsilon (\vec{\mathbf {k}} ) = {1\over {Jn_c r_c^2\rho}} \cdot {1\over {{|} \vec{\mathbf {k}}{ |}^2 + g/(Jn_c r_c^2)}} .
\label{Seps}
\end{equation}
Thus, where $S_{\vec{\mathbf {\pi}}}$ grows without bound as the wavevector $\vec{\mathbf {k}}$ becomes small, $S_\epsilon (\vec{\mathbf {k}} )$ stops growing once $\vec{\mathbf {k}}$ is smaller than a characteristic scale $k_c = 1/\xi = \sqrt{g/(Jn_c r_c^2)}$.  We note that $\xi$ has the dimensions of a length, and we expect that this will set the scale over which correlations extend.  Indeed, if we transform back to get the correlation function, we have
\begin{eqnarray}
C_\epsilon (\vec{\mathbf{x}}) &=&  \int  {{d^3 k}\over{(2\pi)^3}} \, e^{-i\vec{\mathbf {k}}\cdot \vec{\mathbf {x}}} S_\epsilon (\vec{\mathbf {k}} )\\
&=& {1\over {Jn_c r_c^2\rho}} \int  {{d^3 k}\over{(2\pi)^3}} \, e^{-i\vec{\mathbf {k}}\cdot \vec{\mathbf {x}}} 
{1\over {{|} \vec{\mathbf {k}}{ |}^2 + g/(Jn_c r_c^2)}}\nonumber\\
&&\\
&\propto& e^{- |\vec{\mathbf{x}}|/\xi} ,
\end{eqnarray}
corresponding to Eq (\ref{exp_decay}) of the main text.

From these results we can see that, for generic values of $g/J$, the maximum entropy model predicts very different kinds of correlations for directions and speeds. In the case of directions, the correlations have a dominant contribution from long wavelength modes, there is no intrinsic length scale, and we see  scale--free behavior.   On the contrary, in the case of speed fluctuations  the contribution of the long wavelength modes is cut off by the `mass' term (by analogy with field theory \cite{stat-mech}) $g/J$, resulting in correlations that decay exponentially with the distance between birds.   However, when $g/J$ goes to zero, or, more precisely, when the predicted correlation length $\xi$ becomes comparable to the linear dimensions of the flock as whole, our analysis breaks down.  We have described an essentially infinite system, with no boundaries.  When $g/Jn_c$ is small enough that $\xi \sim r_c \sqrt{Jn_c/g} \sim L$, then the whole flock is effective correlated, and a more detailed analysis is needed.  We shall see that, in this ``critical'' regime, it is possible for the speed fluctuations also to be scale--free.

\section{Decoupling speeds and flight directions}
\label{app:decoupling}

The approach we have taken thus far is to build the least structured models that are consistent with the observed similarity of velocities between birds and their near neighbors.  Importantly, we treat the velocities as vectors, and use a measure of similarity that is a rotationally invariant, analytic function of these vectors, $Q_{\rm int}$ in Eqs (\ref{Qint}) and (\ref{app_Qint}).  One could imagine, however, that real birds do not obey these symmetries.  In particular, they could have very separate mechanisms for adjusting their speeds and directions in relation to those of their neighbors, or their perceptual apparatus for estimating speeds and directions may introduce errors that are not equivalent to an isotropic vector error.  Under these conditions, it would make more sense to build models that have separate constraints for the observed degree of speed and direction similarity among neighbors, and this is what we explore in this Appendix.

We can measure the degree of similarity or correlation among directions in the same way that we did in Ref \cite{bialek+al_12}, defining
\begin{equation}
C_{\rm int} = {1\over N}\sum_{{\rm i}=1}^N {1\over {n_c^{\rm dir}}} \sum_{{\rm j}\in {\cal N}_{\rm i}^{\rm dir}} \vec{\bf s}_{\rm i} \cdot \vec{\bf s}_{\rm j} ,
\end{equation}
where we allow that the neighborhood for measuring directional similarity may have a size $n_c^{\rm dir}$ that differs from the corresponding neighborhood for measuring speed similarity, $n_c^{\rm sp}$.  We can also define a (dis)similarity measure for the speeds, by analogy with $Q_{\rm int}$,
\begin{equation}
Q_{\rm int}^{\rm sp} = {1\over {2N}}\sum_{{\rm i}=1}^N {1\over {n_c^{\rm sp}}} \sum_{{\rm j}\in {\cal N}_{\rm i}^{\rm sp}} (v_{\rm i} -  v_{\rm j})^2 .
\end{equation}
If we build the maximum entropy model consistent with measured values of these quantities, plus the mean and variance of individual speeds across the flock, we obtain, instead of  Eq (\ref{hamilton}),
\bea
&&{\cal H}({\bf v}) = \frac{J^{\rm sp}}{4v_0^2} \sum_{{\rm i,j}=1}^N n_{\rm ij}^{\rm sp}  (v_{\rm i} - v_{\rm j})^2 - \frac{J^{\rm dir}}{2v_0^2}\sum_{{\rm i,j}=1}^N n_{\rm ij}^{\rm dir}  \vec{\bf s}_{\rm i}\cdot \vec{\bf s}_{\rm j} 
\nonumber\\
&&\phantom{ppppppppppppp} +\frac{g}{2v_0^2}\sum_{{\rm i}=1}^N v_{\rm i}^2 - {\mu\over{v_0}} \sum_{{\rm i}=1}^N v_{\rm i} ,
\label{H_decouple}
\eea
where $n_{\rm ij}^{\rm sp}$ is defined as with $n_{\rm ij}$ above, but with neighborhoods of size $n_c^{\rm sp}$, and similarly for $n_{\rm ij}^{\rm dir}$.  Notice that we now have two different coupling strengths, $J^{\rm sp}$ and $J^{\rm dir}$, controlling speed and directional ordering, respectively.

Because our original model breaks into separate pieces for directional and speed fluctuations, we can carry over all the calculations, being careful about the values of the parameters.  If we set $J^{\rm sp} = J^{\rm dir}$ we are back to our original model.    With the two separate parameters we find the log--likelihood, by analogy with Eq (\ref{likelihood_border}),
\begin{widetext}
\be
\Phi 
=
-\ln{\cal Z}_{\rm dir} (J^{\rm dir};  n_c^{\rm dir})-\ln {\cal Z}_{\rm sp} (J^{\rm sp}, g; n_c^{\rm sp}) +{\cal H}_{\cal B} - \frac{J^{\rm sp} n_c^{\rm sp} N}{2} \langle Q_{\rm int}^{\rm sp}\rangle_{\rm exp} 
+ \frac{J^{\rm dir} n_c^{\rm dir} N}{2} \langle C_{\rm int}\rangle_{\rm exp}
- N\frac{g}{2}\langle \sigma^2\rangle_{\rm exp} .
\label{likelihood_decouple}
\ee
\end{widetext}

\begin{figure}[tb] 
 \centering
  \includegraphics[scale=0.22]{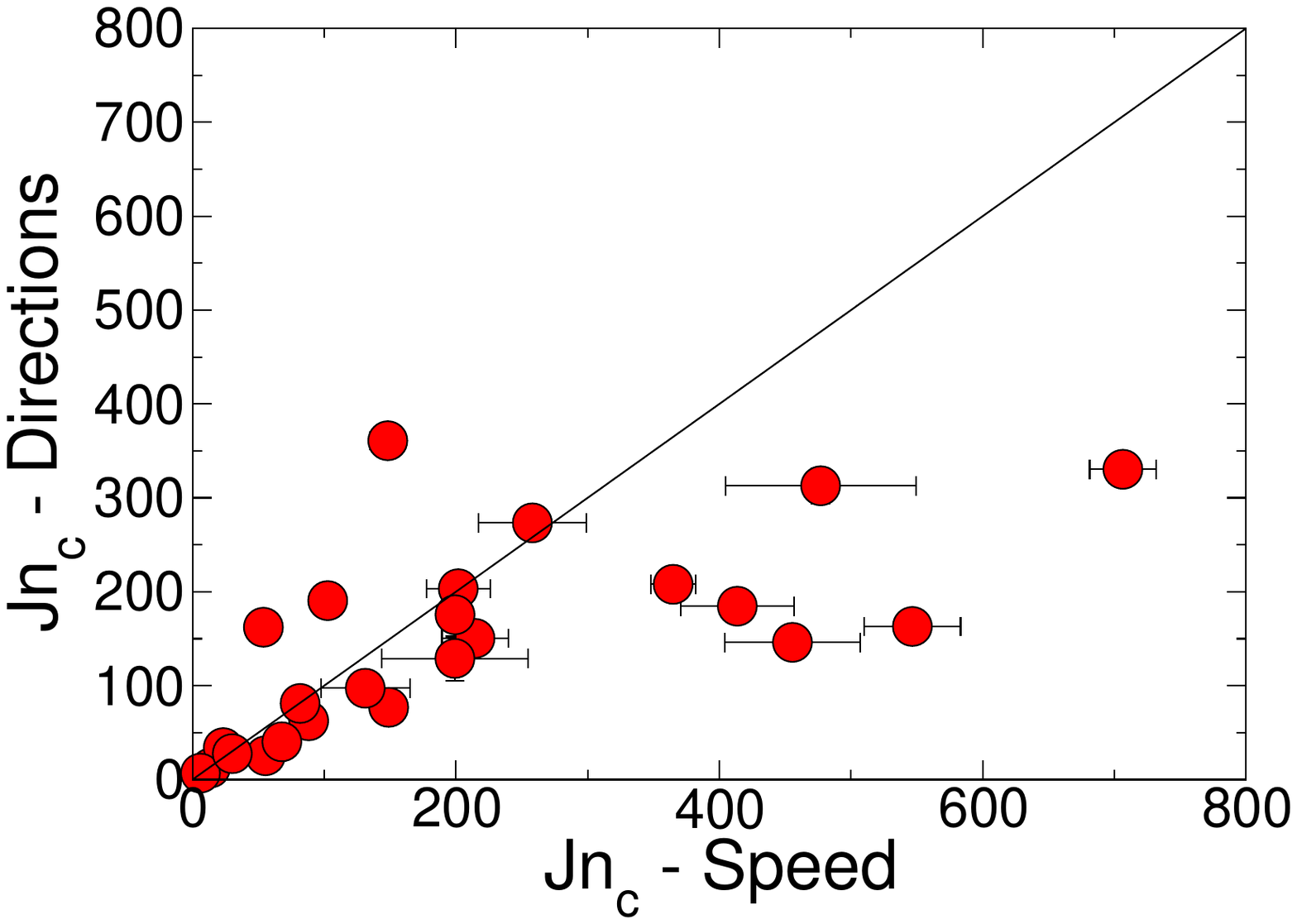}\\ \mbox{(a)}\\\vspace{1 cm}
  \includegraphics[scale=0.22]{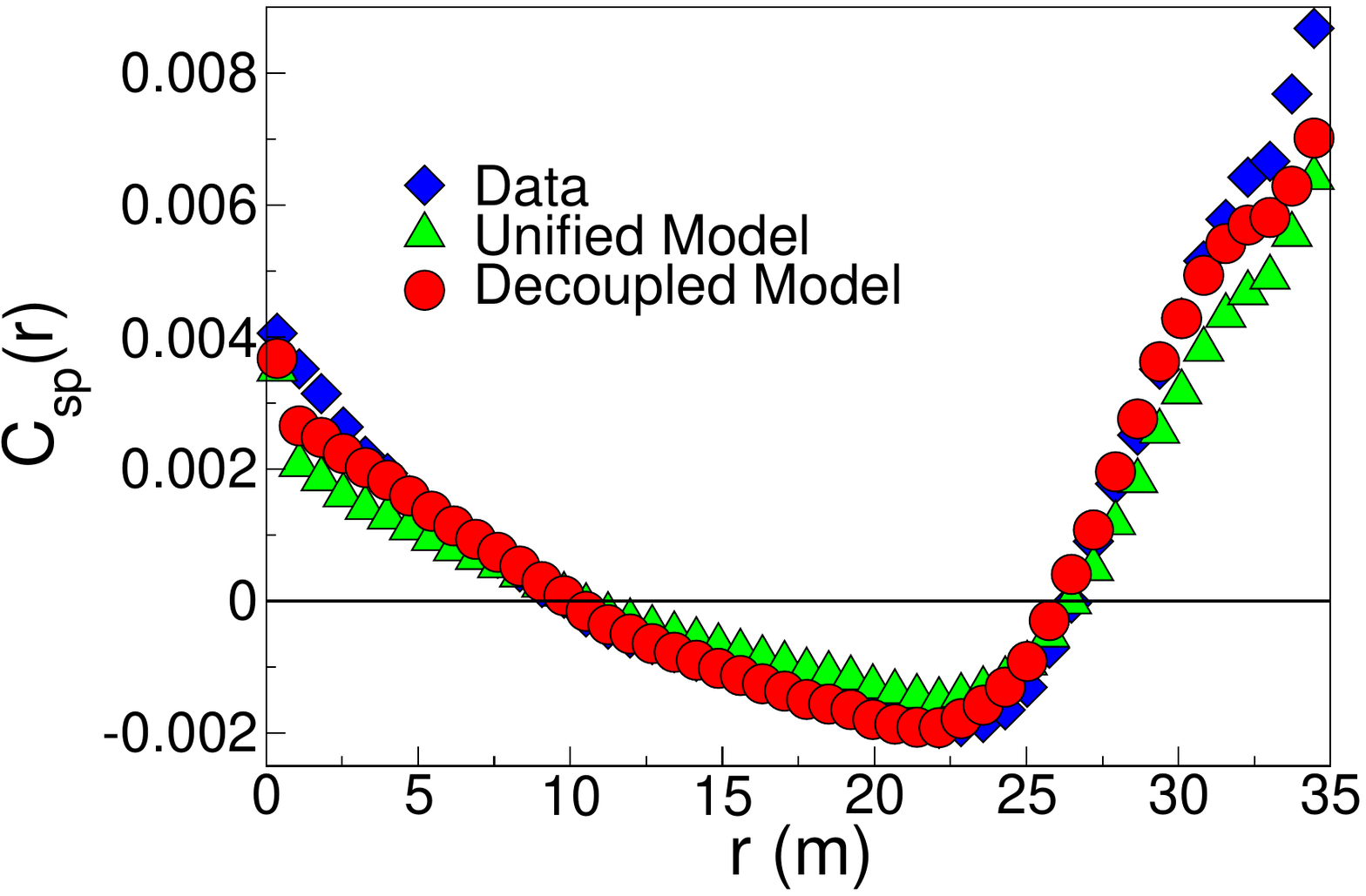}\\ \mbox{(b)}
\caption{Model with independent interactions for speed and flight directions. (a) The inferred global interaction strength $J n_c$ for the orientational degrees of freedom (vertical axis) vs the speed degrees of freedom (horizontal axis). The straight line corresponds to $y=x$, i.e. to the global model where the interaction parameters are the same for speed and flight directions. (b) Prediction for the speed correlation function of the unified model Eq (\ref{hamilton}) and for the decoupled model based on Eq (\ref{H_decouple}), for flock 28-10, corresponding to the point most on the right in panel (a).}
   \label{fig:decoupling}
\end{figure}

We can then infer, independently for speed and orientation, the interaction parameters, and compare them to see how different they are. We can also check whether and how much the predictions for the correlation functions are better than in the simpler, unified model. Results are shown in Fig \ref{fig:decoupling}. We can see that for most flocks the global interaction strength $Jn_c$ for the speed and directional degrees of freedom are very similar to each other (Fig \ref{fig:decoupling}a): in this case the unified model discussed in the previous section is basically equivalent to this more general model, both in terms of values of the inferred parameters and in terms of predictions for the correlation functions. For a few flocks, however, we observe a decoupling between flight directions and speeds. This typically occurs when the fractional speed fluctuations are on a different scale from the directional fluctuations.  In these cases, the model that fixes the local similarities of speed and direction separately provides better predictions for the speed correlations than the unified model (Fig \ref{fig:decoupling}b), although these differences are not huge.

Building a model that fixes the local similarities of speed and direction separately {\em must} provide  a more accurate description of the system, since it imposes two different ways in which our model distribution $P({\bf v})$ has to match the real distribution of (vector) velocities.   The fact that the gain in accuracy usually is small seems significant, and suggests that those rare instances where differences are larger should have biological meaning.  Indeed,  in most of the events where the decoupling is stronger (to the right in Fig \ref{fig:decoupling}a) the flocks are turning. Recent findings \cite{attanasi+al_13} show that additional conservation laws must be taken into account to explain the dynamics during the turn. Even if such conservation laws do not modify the form of the  probability distribution we are investigating in the present work, they might give rise to different effective parameters for directions and speeds.

\section{Dynamical model}
\label{app:dynamics}

In this section we describe more in detail the dynamical model introduced in   Eqs (\ref{dynamics}) and (\ref{kinetic}), and its numerical implementation.  
We have
\bea
\gamma \frac{d\vec{\mathbf{v}}_{\rm i}(t)}{dt}
&=&
-\nabla_{\rm i} {\cal H}(\{\vec{\mathbf{v}}_{\rm j}\}) +\vec{\mathbf{\eta}}_{\rm i} (t) 
\label{dynamics_app}\\
&=& - {J\over{2 v_0^2}} \sum_{\rm j} n_{\rm ij} (\vec{\bf v}_{\rm i} - \vec{\bf v}_{\rm j}) - {g\over{v_0^2}}\frac{\vec{\bf v}_{\rm i}}{v_{\rm i}}\left(v_{\rm i}-\hat{v}\right)
\nonumber\\
&&\,\,\,\,\,\,\,\,\,\,\,\,\,\, + {1\over {n_c}}\sum_{{\rm j}\in {\cal N}_{\rm i}} \vec{\bf f}_{\rm ij} + \vec{\bf \eta}_{\rm i}(t)\label{app_dyn1}\\
\frac{d\vec{\mathbf{x}}_{\rm i}}{dt}&=&\vec{\mathbf{v}}_{\rm i}\label{app_dyn2} ,
\eea
where we have added, as described in the text, forces $\vec{\bf f}_{\rm ij}$ that serve to hold the flock together.  If we write the vector components of $\vec{\bf \eta}_{\rm i}(t)$ as ${\bf \eta}_{\rm i}^{\nu}(t)$, with $\nu = 1, \, 2, \, 3$, then 
\begin{equation}
\langle {\bf \eta}_{\rm i}^{\nu}(t){\bf \eta}_{\rm j}^{\mu}(t')\rangle = 2 \gamma T \delta_{\rm ij} \delta_{\mu\nu}  \delta (t - t'),
\end{equation}
where $T$ is an effective temperature for the noisy dynamics.  We can chose our units of time so that $\gamma = 1$, and from the discussion in   Appendix \ref{app:solving}, we can chose $\hat v = v_0 = \langle V\rangle_{\rm  exp}$, the desired mean speed of the flock.

In this form, the  model that we  are considering describes ``self--propelled particles'' (SPP), and is very similar to the Vicsek model with attraction, which has been studied extensively in the literature \cite{vicsek+al_95,gregoire+al_03,gregoire+chate_04,camperi+al_12}. An attraction term is required to keep the flock cohesive in open space and prevent fluctuations and/or perturbations to disrupt the group. It has been shown that these effects are remarkably less important in models with topological interactions \cite{ballerini+al_08a,camperi+al_12,ginelli+al_10}, which are much more robust in cohesion than SPP models with metric interactions. Nevertheless, even in the topological case, an attraction force is the most controlled way to fix the density of the group to a stationary value, therefore we will include it.
We choose the forces
\be
 \vec{\bf f}_{\rm ij}=\alpha \frac{\vec{\bf r}_{\rm ij}}{r_{\rm ij}}\left\{ 
\begin{array}{lll}
\frac{1}{4} \frac{r_{\rm ij}-r_e}{r_a-r_{hc}} & {\rm if} & r_{\rm ij}<r_a\\
1 & & {\rm otherwise} ,
\end{array}
\right.
\ee
where $\vec{\bf r}_{\rm ij}$ is the vector from bird $\rm i$ to bird $\rm j$, $r_{\rm ij} = |\vec{\bf r}_{\rm ij}|$ is its length; $r_e$ is the equilibrium distance between birds where the force vanishes, while $r_a$ and $r_{hc}$ set spatial scales for the extent of the force.    In our simulations we choose $r_e = 0.5$, $r_a = 0.8$, and $r_{hc} = 0.2$, which sets our units of length, and $\alpha = 0.95$.

An important point is that, when we sum the contributions of the forces $\vec{\bf f}_{\rm ij}$, we include only birds within a limited neighborhood, ${\rm j} \in {\cal N}_{\rm i}$.  As in the measure of similarity $Q_{\rm in}$, this neighborhood is defined topologically, so that each bird feels the  effect of $n_c$ closest neighbors, rather than all the birds within a fixed physical distance.   In addition, for these simulations we introduced a balancing criterion, according to which a bird considers interacting neighbors homogeneously around it to coordinate with. This mimics the idea of a shell of relevant topological neighbors, and is similar to using Voronoi neighbors, as in Ref \cite{ginelli+al_10}, but is much easier to implement numerically.   A balanced interaction enhances the stability of the flock \cite{camperi+al_12}, increasing the range of parameters where Eqs (\ref{app_dyn1}) and  (\ref{app_dyn2}) give rise to realistic behavior. However, we checked also the simple topological case, obtaining qualitatively similar results.

Despite its similarity with other SPP models, the model we are considering has a crucial new ingredient, namely that the speeds of the individual birds are not fixed but can change in time. Accordingly,  Eq (\ref{app_dyn1}) describes the evolution of the full velocity (rather than the flight direction, as in Ref \cite{vicsek+al_95}), with a  term  $\propto g$ that sets the scale of the speed fluctuations. In addition, existing SPP models are usually defined as discrete dynamical update equations, which do not have a well defined continuum limit.  In contrast, we have defined our model as a stochastic differential equation.

We simulate our model using a finite interval (Euler) discretization, and we checked that macroscopic properties of the flock (e.g., the mean speed) remained the same if the size of the time step was decreased.  Parameters $J$ and $n_c$ can be taken from the discussion of real flocks, and the temperature $T$ adjusted until the polarization is in the range seen in the data (Table \ref{table}).  We simulated flocks of different sizes, and checked that the flock had come to a stationary state before making measurements.  With all other parameters fixed, we varied $g$, with the results shown in Fig \ref{fig:dynamics}.

Long ranged correlations can arise through one other mechanism that we have not discussed, and this is the emergence of ``hydrodynamic modes;'' it has been argued that such modes are an essential feature of self--propelled particle models on the largest spatial and temporal scales \cite{toner+tu_95,toner+tu_98}.  The simulations described here suggest, however, that such effects become dominant only on much larger scales in space and especially in time, and thus cannot explain the scale free speed correlations that we observe at equal times in real flocks.  We know that both metric and topological SPP/Vicsek models exhibit giant density fluctuations on large scales \cite{ginelli+al_10}, yet we have seen that as long as $g$ is finite, speed correlations are short range and a critical value of $g$ is necessary to make them scale--free.

\end{document}